\documentclass[useAMS,usenatbib,a4]{mn2e}

\usepackage{graphicx}
\usepackage{amssymb}
\usepackage{amsmath}
\usepackage{datetime}
\usepackage[normalem]{ulem}
\usepackage{times}
\usepackage{hyperref}
\usepackage{booktabs}

\newcommand{\Ro}{\ensuremath{{R_{\mathrm o}}}}
\newcommand{\taustar}{\ensuremath{\tau_{\ast}}}
\newcommand{\Mdot}{\ensuremath{\rm \dot{M}}}
\newcommand{\xspec}{{\sc xspec}}
\newcommand{\vinf}{\ensuremath{v_{\infty}}}
\newcommand{\chandra}{\textit{Chandra}}
\newcommand{\xmm}{\textit{XMM}}
\newcommand{\Rstar}{\ensuremath{R_{\ast}}}

\newcommand{\Msunyr}{\ensuremath{{\mathrm {M_{\sun}~{\mathrm yr^{-1}}}}}}

\newcommand{\zpup}{$\zeta$ Pup}

\newcommand{\hetgs}{HETGS}

\newcommand{\beq}{\begin{equation}}
\newcommand{\eeq}{\end{equation}}
\newcommand{\beqa}{\begin{eqnarray}}
\newcommand{\eeqa}{\end{eqnarray}}

\usepackage{xcolor}
\usepackage{soul}
\definecolor{david}{rgb}{0.25, 0.5, 0.99}

\definecolor{old}{rgb}{0.75, 0.75, 0.75}

\definecolor{lamiaa}{rgb}{0.55, 0.71, 0}

\definecolor{victoria}{rgb}{0.95, 0.55, 0.8}

\definecolor{graham}{rgb}{0.90, 0.5, 0.3}

\definecolor{maurice}{rgb}{0.70, 0.1, 0.9}

\DeclareGraphicsExtensions{.pdf,.png,.jpg}

\begin{document}

\title[\zpup\/ \chandra\/ line profile changes]{\chandra\/ spectral measurements of the O supergiant $\zeta$ Puppis indicate a surprising increase in the wind mass-loss rate over 18 years}
\author[D.Cohen et al.]{David H.\ Cohen,$^{1}$\thanks{E-mail:
    dcohen1@swarthmore.edu} Jiaming Wang,$^{1}$ V\'{e}ronique Petit,$^{2}$ Maurice A. Leutenegger,$^{3}$ 
\newauthor Lamiaa Dakir,$^{4}$ Chloe Mayhue,$^{1}$ Alexandre David-Uraz$^{2}$
\\
  $^{1}$Swarthmore College, Department of Physics and Astronomy, Swarthmore, Pennsylvania 19081, USA\\
  $^{2}$University of Delaware, Department of Physics and Astronomy  \& Bartol Research Institute, Newark, Delaware 19716, USA \\
  $^{3}$NASA/Goddard Space Flight Center, Code 662, Greenbelt, Maryland 20771, USA \\
  $^{4}$Bryn Mawr College, Bryn Mawr, Pennsylvania 19010, USA \\
}

\maketitle

\label{firstpage}

\begin{abstract}
New long \chandra\/ grating observations of the O supergiant \zpup\/ show not only a brightening of the x-ray emission line flux of 13 per cent in the 18 years since {\chandra}'s first observing cycle, but also clear evidence -- at more than four sigma significance -- 
of increased wind absorption signatures in its Doppler-broadened line profiles. We demonstrate this with non-parametric analysis of the profiles as well as Gaussian fitting and then use line-profile model fitting to derive a mass-loss rate of $2.47 \pm 0.09 \times 10^{-6}$ \Msunyr, which is a 40 per cent increase over the value obtained from the cycle 1 data. The increase in the individual emission line fluxes is greater for short-wavelength lines than long-wavelength lines, as would be expected if a uniform increase in line emission is accompanied by an increase in the wavelength-dependent absorption by the cold wind in which the shock-heated plasma is embedded. \end{abstract}


\section{Introduction}

Resolved x-ray emission line profiles provide diagnostic information about both the x-ray production in the dense, highly supersonic radiation-driven winds of O stars, and also about the mass-loss rates of these winds.  Indeed, one of the first results for massive stars provided by the \chandra\/ spectrometers soon after the observatory's launch in 1999 was the confirmation that these x-rays arise in the stellar wind, rather than in a magnetically confined corona, as is the case for low-mass stars. This was revealed by the very significant Doppler broadening (half-width at half-maximum $ \approx 1000$ km s$^{-1}$) in the x-ray emission line profiles of the canonical single O supergiant, \zpup\/ \citep{Kahn2001, Cassinelli2001}. 

In addition to verifying the wind origin of the x-rays in O stars, measuring resolved x-ray line profiles enables us to constrain the spatial distribution of the wind-shocked plasma, as there is a mapping between distance from the photosphere and wind speed. Analysis of the cycle 1 \chandra\/ grating spectra of \zpup\/ (taken in 2000) showed that the x-ray emission begins about half a stellar radius into the wind flow, as expected for embedded wind shocks produced by the line-deshadowing instability \citep{Cohen2010}, and this result holds for other O stars with strong winds observed by \chandra\/ \citep{Cohen2014a}.  

The line profiles are also affected by continuum absorption in the cold component of the wind, primarily from K-shell photoionization of metals. This attenuation preferentially affects the rear, red-shifted hemisphere of the wind, leading to characteristically blue-shifted and asymmetric profiles, with the degree of asymmetry being governed by the wind column density and hence, mass-loss rate \citep{MacFarlane1991,Ignace2001,OC2001}. \citet{OC2001} presented a line profile model with three free parameters: line flux, x-ray onset radius (\Ro), and characteristic wind optical depth (\taustar), which can be fit to individual resolved x-ray lines. The optical depth parameter, \taustar, is a function of wavelength via the wavelength dependence of bound-free continuum opacity in the cool wind, and so every line in a given spectrum is expected to have an optical depth proportional to the wind opacity at the wavelength of that line, and the ensemble of fitted \taustar\/ values can be used to derive a wind mass-loss rate. This procedure was used to fit sixteen lines and line complexes in the cycle 1 \chandra\/ spectrum of \zpup, and it was found that a single x-ray onset radius of $\Ro = 1.5$ \Rstar\/ is consistent with the fitting results of all the lines, while the individual line's fitted \taustar\/ values gave a mass-loss rate of $1.76 ^{+0.13}_{-0.12} \times 10^{-6}$ \Msunyr\/ \citep{Cohen2014a}. 

X-ray emission from embedded wind shocks (EWS) in O stars is generally not variable on short timescales comparable to wind flow and shock cooling times (hours), which is taken as an indication that a very large number of wind shock zones contribute to the overall x-ray emission \citep{NOG2013}. Wind properties of O stars, including \zpup, often show cyclical short term variability \citep{Massa1995, Howarth1995}, but the global wind properties -- including the mass-loss rate -- seem to be quite constant in normal O stars. However, a long x-ray observing campaign with \xmm\/ showed that the overall x-ray emission levels of \zpup\/ are variable with an amplitude of variability of about 20 percent, but no clear timescale of variability\footnote{Due to the relatively large wavelength calibration uncertainties of the \xmm\/ Reflection Grating Spectrometer (RGS), it is not possible to ascertain whether the line profiles vary from observation to observation.}, though longer than $\sim 10^5$ s \citep{NOG2013}. 

During cycle 19 \chandra\/ carried out a long sequence of 21 observations of $\zeta$ Puppis with the High Energy Transmission Grating Spectrometer (HETGS). In this paper we present results from the analysis of the ten measurable lines and line-complexes in these data and compare them to those previously obtained from the cycle 1 dataset, taken 18 years earlier, specifically focusing on the wind optical depths and mass-loss rate. 

In \S2 we describe the data, its reduction, and the three different approaches we use for analyzing the line profile shapes. In \S3 we present the results of the line profile fitting. In \S4 we discuss the modeling results, including changes in the x-ray and wind properties between the two sets of observations, which imply a large change in the mass-loss rate, and in \S5 we summarize our conclusions.

\section{Data, analysis, and modeling}

\begin{table}
\begin{center}
  \caption{$\chandra$ Observing Log}
\begin{tabular}{cccc}
  \hline
 Observation ID & Exposure time & Date & Cycle number \\
  & (ks) & & \\
  \hline
   640  & 67.74  &  2000 Mar 28 & 1  \\
  21113 & 17.72  &  2018 Jul 1 & 19  \\
  21112 & 29.70  &  2018 Jul 2 & 19  \\
  20156 & 15.51  &  2018 Jul 3 & 19  \\
  21114 & 19.69  &  2018 Jul 5 & 19  \\
  21111 & 26.86  &  2018 Jul 6 & 19  \\
  21115 & 18.09  &  2018 Jul 7 & 19  \\
  21116 & 43.39  &  2018 Jul 8 & 19  \\
  20158 & 18.41  &  2018 Jul 30 & 19  \\
  21661	& 96.88	 &  2018 Aug 3 & 19  \\
  20157	& 76.43  &  2018 Aug 8 & 19  \\
  21659	& 86.35	 &  2018 Aug 22 & 19 \\
  21673	& 14.95	 &  2018 Aug 24 & 19 \\
  20154	& 46.97	 &  2019 Jan 25 & 19 \\
  22049	& 27.69	 &  2019 Feb 1 & 19 \\
  20155	& 19.69	 &  2019 Jul 15 & 19 \\
  22278	& 30.51	 &  2019 Jul 16 & 19 \\
  22279	& 26.05	 &  2019 Jul 17 & 19 \\
  22280	& 25.53	 &  2019 Jul 20 & 19 \\
  22281	& 41.74	 &  2019 Jul 21 & 19 \\
  22076	& 75.12	 &  2019 Aug 1 & 19 \\
  21898	& 55.70	 &  2019 Aug 17 & 19 \\
  \hline
\end{tabular}
\end{center}
\label{tab:observing_log}
\end{table} 

\begin{figure*}
	\includegraphics[angle=0,width=0.48\textwidth]{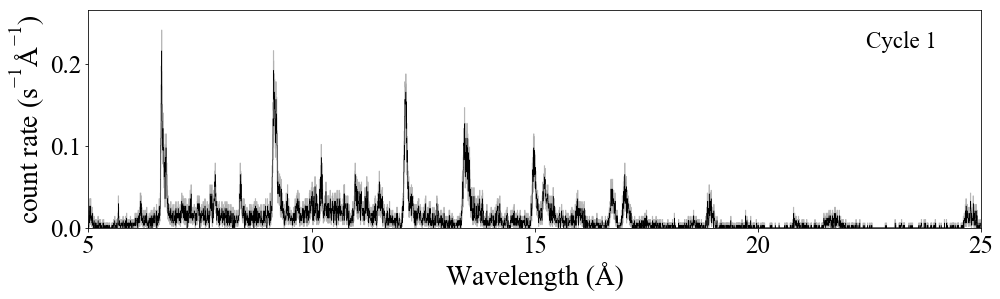}
	\includegraphics[angle=0,width=0.48\textwidth]{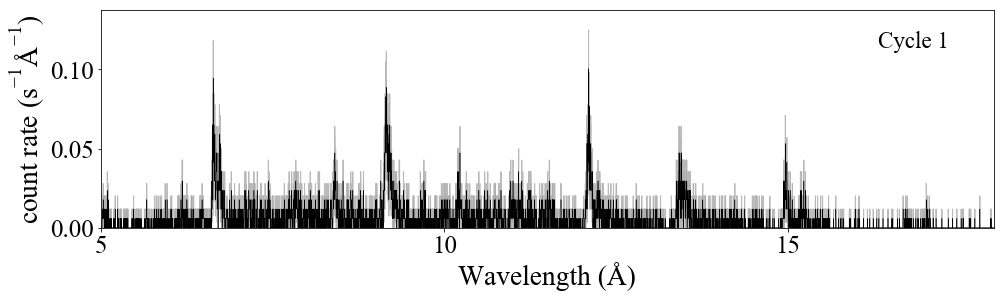}
	
	\includegraphics[angle=0,width=0.48\textwidth]{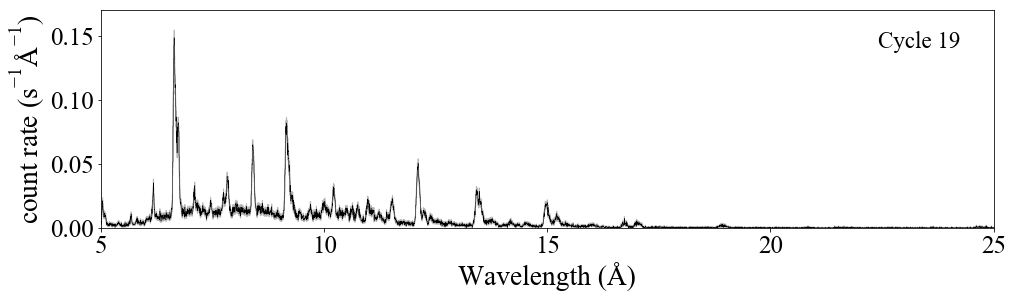}
	\includegraphics[angle=0,width=0.48\textwidth]{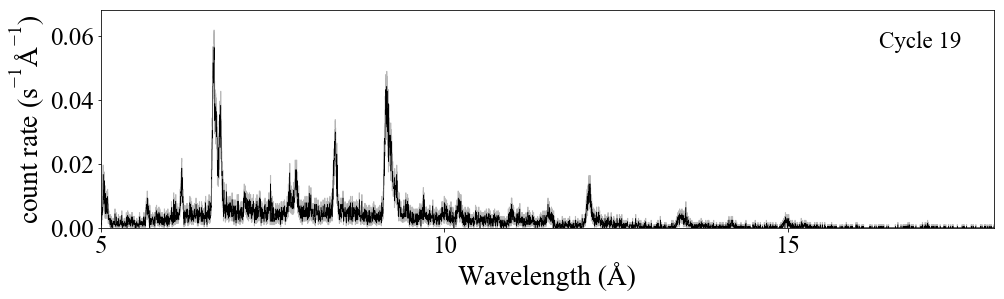}
	
	\includegraphics[angle=0,width=0.48\textwidth]{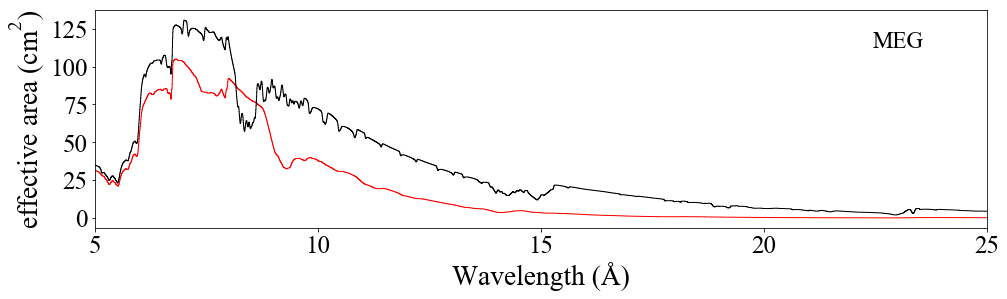}
	\includegraphics[angle=0,width=0.48\textwidth]{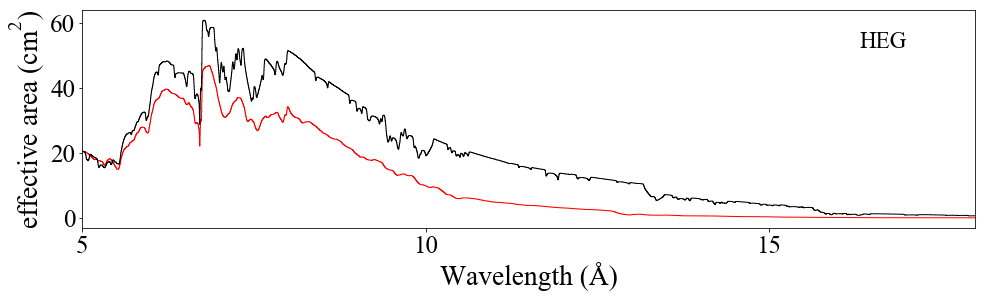}
	\caption{The first-order co-added $\pm 1$ MEG (left) and HEG (right) spectra for cycle 1 and the combined cycle 19 observations are shown in the first two rows. The last row shows the exposure-time-weighted average effective area of the \chandra\/ MEG (left) and HEG (right) at the time of each observation, with the cycle 1 effective in black and the cycle 19 effective area in red. Although count rates are generally lower in the cycle 19 data, modeling we present in \S3 shows that the x-ray flux of \zpup\/ has actually increased between cycle 1 and cycle 19.
	}
	\label{fig:data}
\end{figure*}

Nearly 813 ks of new \chandra\/ observations (PI: W. Waldron) were obtained during cycle 19, between July 2018 and August 2019, in 21 separate pointings (designated by a unique {\it Observation ID}, or Obs ID) ranging in exposure time from just over 10 ks to almost 100 ks. We present an observing log in Table \ref{tab:observing_log}, that also includes the lone Obs ID from cycle 1, taken 18 years earlier. 

The \chandra\/ \hetgs\/ contains two grating arrays -- the Medium Energy Grating (MEG) and High Energy Grating (HEG) \citep{Canizares2005}. We reprocessed the data retrieved from the archive in November 2019 using the CIAO pipeline (v.\ 4.11), including CALDB (v.\ 4.8.5), producing response matrix files and auxiliary response files (containing effective area information), along with extracted first-order spectra for both the MEG and HEG. We do not subtract a background when we analyze the data. 

We show the MEG and HEG spectra for all the cycle 19 observations, combined, in Fig.\ \ref{fig:data}. Note that this is for illustration only. Inspecting the combined 21 Obs IDs we identify seven individual lines and three He-like line complexes that can be analyzed. Due to the deterioration of the long-wavelength sensitivity there are several lines (and line complexes) seen in the cycle 1 data which we do not detect in the cycle 19 data, even when all Obs IDs are combined. 

All the line analysis we present here is performed on the 21 separate data sets, with each dataset consisting of a co-added ($\pm 1$ order) MEG spectrum and a co-added ($\pm 1$ order) HEG spectrum. For each line or line complex we analyze, we treat these 42 spectra as a single data set. We do not combine them, rather we analyze them simultaneously. We use \xspec\/ v.\ 12.9 to fit a flat spectral model to carefully selected continuum regions on either side of each of the lines and line complexes. We use this continuum modeling for all of the analyses presented in this section, and in Appendix \ref{app:continuum} we provide details, including figures, describing the continuum fits.  

Prior to fitting the wind-profile model from which wind optical depths and the wind mass-loss rate can be measured, we analyze the seven single lines in the spectra in two model-independent ways: (1) a moment analysis that treats each line profile as a probability distribution \citep{Cohen2006} and (2) fitting Gaussian profiles.   

\begin{figure}
\begin{center}
	\includegraphics[angle=0,width=0.48\textwidth]{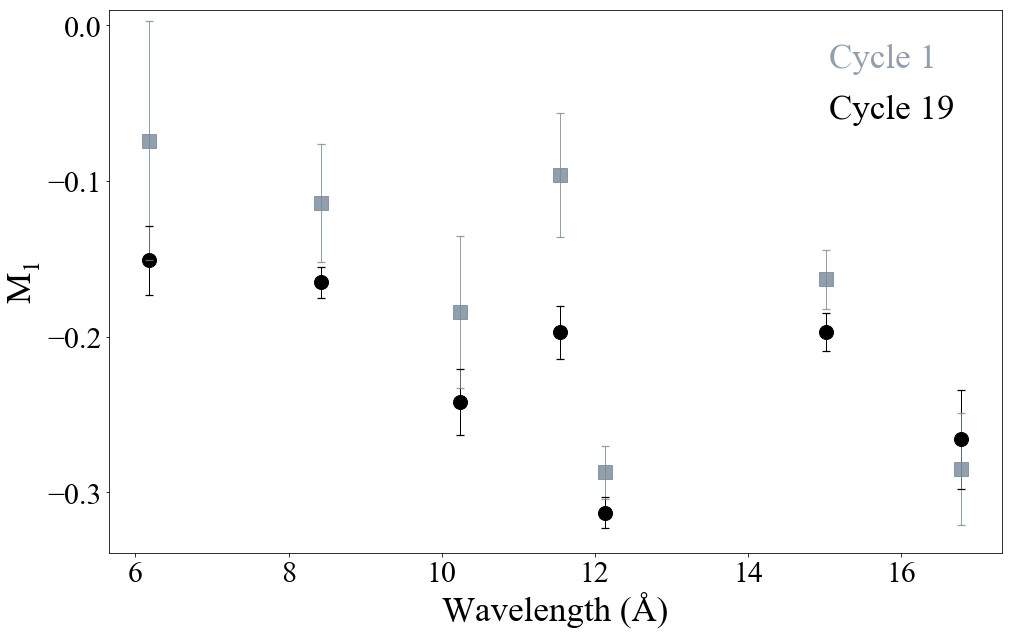}
\end{center}
	\caption{The first moment of each unblended line profile can be seen to change systematically between cycle 1 and cycle 19. 
	}
	\label{fig:moments}
\end{figure}

For the moment analysis, we co-add the 21 Obs IDs, subtract the continuum model from the line, convert the wavelength scale to a unitless Doppler shift scaled to the wind terminal velocity, $x \equiv (\frac{\lambda}{{\lambda}_{\rm o}} - 1)\frac{c}{v_{\rm \infty}}$, and compute the first moment -- the centroid shift of each line according to:

$$
M_{\rm 1} \equiv \frac{{\Sigma}_{i = 1}^{N} x_{i} f(x_i)}{{\Sigma}_{i = 1}^{N} f(x_{i})},
$$

\noindent
where $f$ is the count rate in each bin and sums are over the $N$ data points on $x = [-1:1]$. We do this for each of the seven individual lines (not the heavily blended He-like complexes), analyzing the MEG and HEG data separately, and present the weighted average of the MEG and HEG results for each line in Tab.\ \ref{tab:momGau} and Fig.\ \ref{fig:moments}. 

\begin{figure}
\begin{center}
	\includegraphics[angle=0,width=0.48\textwidth]{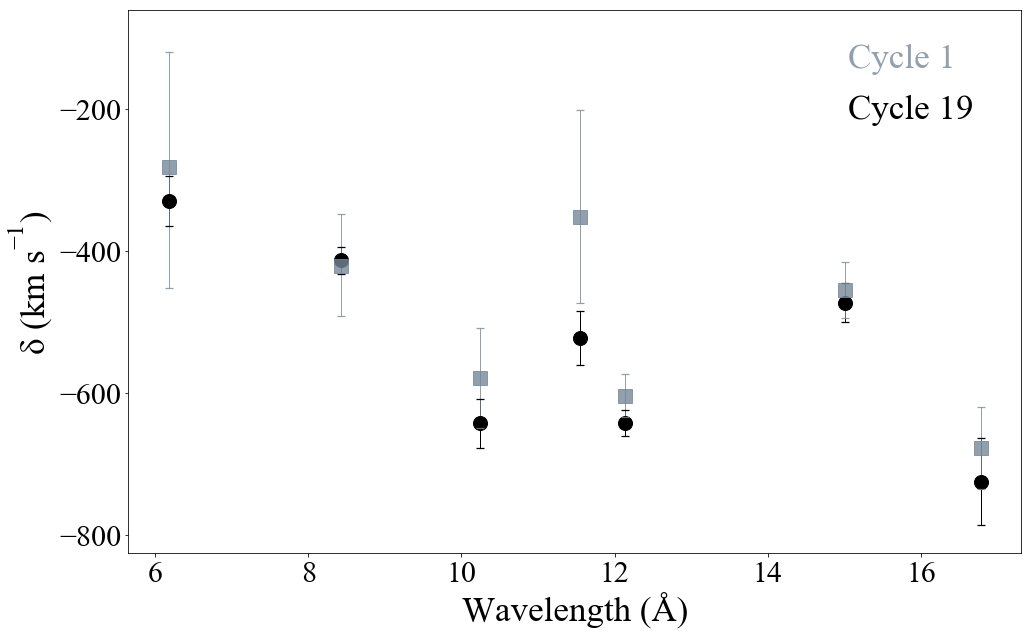}
	\includegraphics[angle=0,width=0.48\textwidth]{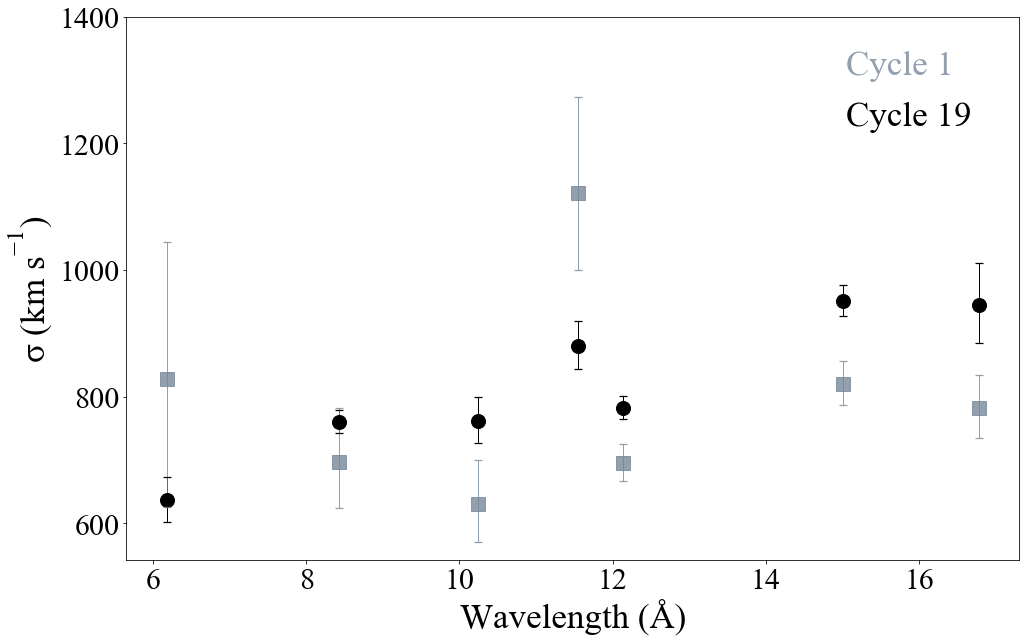}
\end{center}
	\caption{The centroid (top) and width (bottom) of the Gaussian fit to each unblended line profile change systematically between cycle 1 and cycle 19.}
	\label{fig:Gaussian}
\end{figure}

For the Gaussian fitting, we fit a Gaussian line profile model on top of a power-law continuum model, with the continuum model's flux level fixed (at the same level used for the moment analysis and the wind profile fitting). We allow the normalization, centroid, and width of the Gaussian to be free parameters and show the results for the centroid and width, both in km s$^{-1}$, in Tab.\ \ref{tab:momGau} and Fig.\ \ref{fig:Gaussian}. 

\begin{table*}
  \caption{Line profile moment and Gaussian fit results of emission line profiles for cycle 1 and cycle 19}
\begin{tabular}{cccccccc}
  \hline
ion & wavelength & \ $M_{1}$ \ & \ $M_{1}$ \ & centroid shift & centroid shift & width & width\\
  & (\AA) & (cycle 1) & (cycle 19) & (cycle 1, km s$^{-1}$) & (cycle 19, km s$^{-1}$) & (cycle 1, km s$^{-1}$) & (cycle 19, km s$^{-1}$)\\
  \hline
Si {\sc xiv}  & 6.18  & $-0.074 \pm 0.077$ & $-0.151 \pm 0.022$ & $-281_{-170}^{+161}$ & $-330_{-36}^{+35}$ & $827_{-202}^{+217}$ & $636_{-35}^{+36}$ \\
Mg {\sc xii}  & 8.42  & $-0.114 \pm 0.038$ & $-0.165 \pm 0.010$ & $-421_{-71}^{+73}$ & $-414_{-20}^{+20}$ & $697_{-73}^{+85}$ & $760_{-18}^{+19}$ \\
Ne {\sc x}  & 10.24  & $-0.184 \pm 0.049$ & $-0.242 \pm 0.021$ & $-579_{-71}^{+70}$ & $-643_{-35}^{+35}$ & $631_{-61}^{+69}$ & $762_{-35}^{+37}$ \\
Ne {\sc ix}  & 11.54 & $-0.096 \pm 0.040$ & $-0.197 \pm 0.017$ & $-353_{-121}^{+151}$ & $-522_{-39}^{+38}$ & $1121_{-121}^{+151}$ & $880_{-37}^{+40}$ \\
Ne {\sc x}  & 12.13  & $-0.287 \pm 0.017$ & $-0.313 \pm 0.010$ & $-605_{-30}^{+31}$ & $-643_{-19}^{+19}$ & $695_{-27}^{+30}$ & $783_{-18}^{+18}$ \\
Fe {\sc xvii}  & 15.01  & $-0.163 \pm 0.019$ & $-0.197 \pm 0.012$ & $-455_{-39}^{+40}$ & $-474_{-27}^{+28}$ & $820_{-33}^{+36}$ & $951_{-24}^{+25}$ \\
Fe {\sc xvii}  & 16.78  & $-0.285 \pm 0.036$ & $-0.266 \pm 0.032$ & $-678_{-58}^{+58}$ & $-726_{-62}^{+61}$ & $782_{-48}^{+52}$ & $945_{-61}^{+66}$ \\
  \hline
\end{tabular}
\label{tab:momGau}
\end{table*} 

Both the first moment and the Gaussian centroid become more negative between cycle 1 and cycle 19 for six of the seven lines, indicating that the emission lines are systematically more blue-shifted, as would be expected from an increase in wind absorption. In the aggregate, this result -- that the first moment has increased -- is significant at the three sigma level. The Gaussian widths also increase significantly between cycles 1 and 19, which is an expected effect of the line profile shape change associated with increased wind absorption \citep{OC2001}.

Given these indications of morphological changes in the profile shapes consistent with the wind optical depths increasing, we next fit the wind profile model that has been used to derive mass-loss rates for O stars from \chandra\/ grating spectra. The characteristic optical depth parameter, 
$$\taustar = \frac{{\kappa_{\rm \lambda}}\Mdot}{4{\pi}\Rstar{\vinf}}
$$

\noindent 
enables the determination of the wind mass-loss rate given a model of the wavelength-dependent wind opacity, $\kappa_{\rm \lambda}$. We use the same Solar metallicity wind opacity model here that was used in \citet{Cohen2014a}.  

We use the {\it windprof} local model\footnote{Local model documentation and code for both {\it windprof} \citep{OC2001} and {\it hewind} \citep{Leutenegger2006} is available at: \url{https://heasarc.gsfc.nasa.gov/xanadu/xspec/models/windprof.html}.} in \xspec\/ and for the helium-like line complexes in the spectra, we use the variant {\it hewind} that implements a superposition of three profile models at the appropriate wavelengths and incorporates alteration of the forbidden-to-intercombination line ratio according to the model's assumed spatial distribution of the x-ray emitting plasma (governed by \Ro). We find best-fit model parameters by minimizing the C statistic, which is necessary for these unbinned data that have many bins with few counts in the line wings \citep{Cash1979}. We then place confidence limits on each of the three model parameters (flux, \Ro, \taustar) using the $\Delta$C formalism \citep{Nousek1989}, one at a time, with the other two parameters free to vary. We use 68 percent confidence limits (${\Delta}C = 1$) for the line fitting results presented here.

\section{Wind profile model fitting results}

The results of fitting each of the ten lines and line complexes are presented in Table \ref{tab:results}. Individual line profile fits are shown in Appendix \ref{app:profiles}. 

\begin{table*}
  \caption{Emission line parameters}
\begin{tabular}{ccccc}
  \hline
ion & wavelength & \taustar\ & \Ro\ & line flux \\
  & (\AA) & & (\Rstar) &  ($10^{-5}$ ph cm$^{-2}$ s$^{-1}$) \\
  \hline
S {\sc xv}  & 5.04, 5.07, 5.10  &  $0.12_{-.08}^{+.11}$ & $1.48_{-.05}^{+.06}$  & $3.28_{-.10}^{+.10}$   \\
Si {\sc xiv}  & 6.18  & $0.55_{-.14}^{+.17}$ & $1.34_{-.05}^{+.05}$ & $1.06_{-.04}^{+.04}$   \\
Si {\sc xiii}  & 6.65, 6.69, 6.74  &  $0.66_{-.05}^{+.06}$ & $1.60_{-.02}^{+.02}$ & $15.4_{-.1}^{+.1}$   \\
Mg {\sc xii}  & 8.42  &  $1.01_{-.0.11}^{+.12}$ & $1.54_{-.04}^{+.04}$ & $3.95_{-.08}^{+.08}$   \\
Mg {\sc xi}  & 9.17, 9.23, 9.31  &  $0.80_{-.07}^{+.07}$ & $1.77_{-.03}^{+.03}$ & $23.5_{-.3}^{+.3}$   \\
Ne {\sc x}  & 10.24  &  $2.59_{-.16}^{+.19}$ & $1.01_{-.01}^{+.41}$ & $4.13_{-.14}^{+.15}$  \\
Ne {\sc ix}  & 11.54 & $1.54_{-.25}^{+.29}$ & $1.73_{-.11}^{+.10}$ & $7.44_{-.24}^{+.25}$  \\
Ne {\sc x}  & 12.13  &  $3.10_{-.12}^{+.13}$ & $1.01_{-.01}^{+.18}$ & $30.1_{-.5}^{+.6}$  \\
Fe {\sc xvii}  & 15.01  & $2.40_{-.30}^{+.34}$ & $1.87_{-.16}^{+.14}$  & $60.2_{-1.5}^{+1.5}$  \\
Fe {\sc xvii}  & 16.78  &  $3.87_{-.65}^{+.73}$ & $1.46_{-.46}^{+.22}$ & $32.4_{-1.8}^{+1.7}$   \\
  \hline
\end{tabular}
\label{tab:results}
\end{table*}  

The optical depth, \taustar, results listed in Table \ref{tab:results} include 68 per cent confidence limits. These confidence limits characterize probability distributions which are sometimes far from Gaussian. We derive a wind mass-loss rate by fitting a model of wavelength-dependent \taustar\/ values to the ensemble of values derived from fitting individual line profiles. Traditionally this fitting has used $\chi^2$ as a fit statistic and assumed that the uncertainties on that fitted model parameter are Gaussian \citep{Cohen2014a}. Here we relax that assumption and use the actual probability distributions derived from the ${\Delta}C$ values (computed via the {\it steppar} command in \xspec) \citep{Cash1979}. We then fit the mass-loss rate to the ensemble of \taustar\/ probability distributions by maximizing the combined probability of the ten \taustar\/ distributions. 

\begin{figure}
\centering
    \includegraphics[angle=0,width=0.48\textwidth]{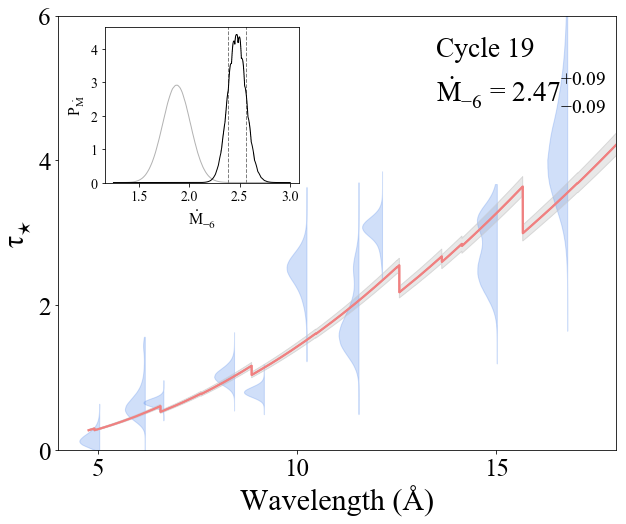}
        \includegraphics[angle=0,width=0.48\textwidth]{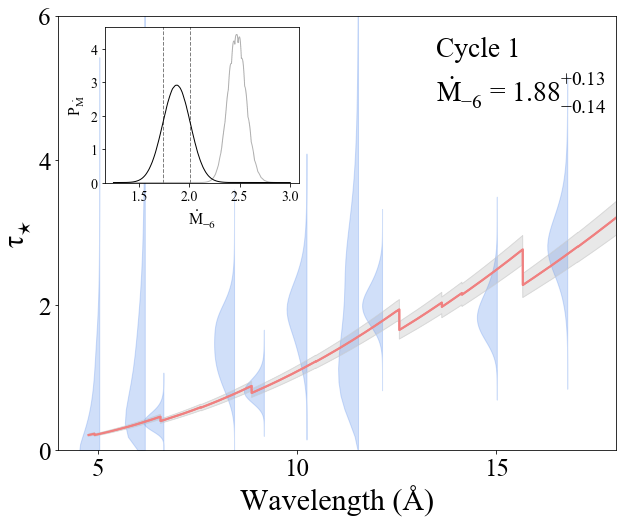}
	\caption{The \taustar\/ probability distribution for each line is shown in light blue, while the best-fit mass-loss rate model is shown in red, with the 68 per cent error band on the mass-loss rate shown in gray surrounding the best-fit model. The cycle 19 results are shown in the top panel, while the reanalyzed cycle 1 results are shown in the lower panel. The overall mass-loss rate probability distributions are shown in the insets, with the 68 per cent confidence limits shown graphically. The blue \taustar\/ probability distributions are truncated at $\pm 5 \sigma$.
	}
\label{fig:Mdot}
\end{figure}

This result is graphically presented in Fig.\ \ref{fig:Mdot} where we show the \taustar\/ probability distributions and the derived mass-loss rate probability distribution itself, from which we find the best-fit mass-loss rate and 68 per cent confidence limits of $2.47 \pm 0.09 \times 10^{-6}$ \Msunyr. This mass-loss rate represents a significant increase -- more than 4 sigma -- over the value derived from the cycle 1 observations, $\Mdot = 1.76^{+0.13}_{-0.12} \times 10^{-6}$ \Msunyr\/ \citep{Cohen2014a}. We note that we re-analyzed the cycle 1 data - including only the same lines analysed in the new cycle 19 data and using this new method, with the fits shown graphically in Appendix \ref{app:profiles}. That reanalysis, shown in Fig.\ \ref{fig:Mdot}, is consistent with the earlier result: $\Mdot = 1.88^{+0.13}_{-0.14} \times 10^{-6}$ \Msunyr. 

\begin{figure}
\begin{center}
	\includegraphics[angle=0,width=0.45\textwidth]{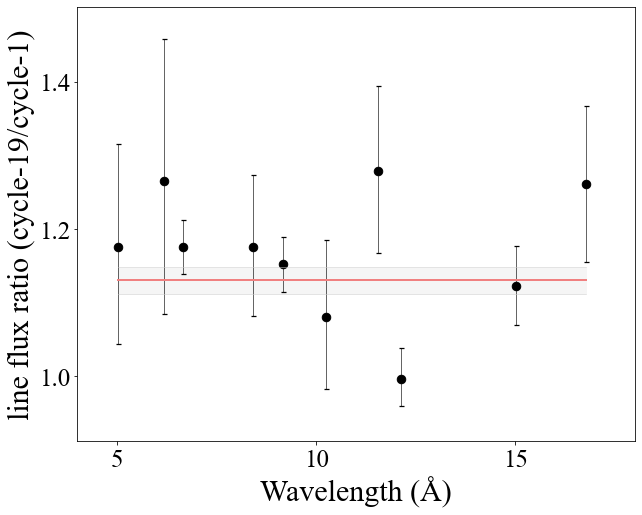}
\end{center}
	\caption{The line flux ratios with 68 per cent confidence limits for all the emission lines measured in both cycle 19 and cycle 1 show an average increase of 13 per cent from cycle 1 to cycle 19 (red line). This increase has some modest wavelength dependence.   
   }
	\label{fig:norms}
\end{figure}

The increase in the wind mass-loss rate is accompanied by a corresponding increase in the emission line fluxes between cycle 1 and cycle 19 (already reported by \citealt{Huenemoerder2020} for short-wavelength lines), as shown in Fig.\ \ref{fig:norms}. This line flux increase averages 13 per cent, but shorter wavelength lines show a consistently larger increase while longer wavelength lines, where the wind opacity is higher, generally show a smaller increase. This is exactly what is expected if all line luminosities increase by the same amount but a corresponding increase in the wind absorption partially compensates for the increase at wavelengths where the wind is optically thick. 

\section{Discussion}

The x-ray luminosity of dense O star winds with radiative shocks is expected to scale linearly with the mass-loss rate \citep{Owocki2013}. So if the mass-loss rate of \zpup\/ has increased by about 40 percent, the x-ray flux should as well, and that is close to what is seen at short wavelengths where the wind is optically thin (see Fig.\ \ref{fig:norms}).  The wind is optically thick to longer wavelength line emission and so the emergent flux of longer wavelength lines will not increase the full 40 percent. This scenario -- a 40 percent increase in wind mass-loss rate between 2000 and 2018 accompanied by roughly the same increase in intrinsic line emission, but because of increased wind absorption, the increase in observed line flux is less than this --  is the simplest interpretation of the observed changes to the x-ray properties of \zpup.  

A mass-loss rate change on years timescale would indeed be quite surprising for a relatively normal O supergiant like \zpup, which is not an LBV or member of any other class of strongly variable evolved massive stars. It is true that \zpup\/ shows small amplitude optical photometric and emission line variability on rotational (days) timescales that implies photospheric hot spots which drive wind variability \citep{Tahina2018}. This observed short timescale periodic variability has a small amplitude and may affect overall mass-loss rates and wind column densities and thus x-ray absorption, but not at the 40 per cent level \citep{David-Uraz2017, Sundqvist2018}. 

The standard line-driven wind theory \citep{CAK} has a scaling of mass-loss rate with bolometric luminosity of roughly $\Mdot \propto L_{\rm bol}^{1.7}$ (see \citealt{OC1999}), suggesting that a brightening of a few tenths of a magnitude may have accompanied a 40 per cent mass-loss rate increase. This is well beyond what is seen in the recent high-cadence photometric monitoring \citep{Tahina2018}. Perhaps the spot characteristics change on longer timescales in ways that affect the global mass-loss rate. However if the global luminosity of the star changes, whether uniformly or via evolving spot characteristics, it would imply a not insignificant internal change to the star. Beating between closely spaced pulsational modes could cause long-timescale brightness variations, though there is no evidence for this particular behavior in \zpup. 

The prior \xmm\/ detection of x-ray variability showed an overall decreasing trend with superimposed stochastic-seeming variability having an amplitude of approximately 20 per cent on days to months timescale \citep{NOG2013}. Interestingly, the x-ray output of the star seems more constant overall in the recent \chandra\/ observations than it was about a decade earlier when the \xmm\/ observations were made. The \xmm\/ light curve implies that the changes to the wind properties of \zpup\/ in the 18 year interval between the two \chandra\/ observing campaigns were neither smooth nor totally abrupt. Broadband spectral trends in the \xmm\/ measurements also indicate that brighter x-ray emission is correlated with a hardening of the spectrum below 1.2 keV \citep{Naze2018}, consistent with the trend we see in the wavelength-dependent line flux changes in the \chandra\/ observations and with the expectations of increased wind column density leading to more soft x-ray attenuation. 

\section{Conclusions}

The x-ray emission lines of \zpup\/ show a systematic increase in blue shift and line shape between the two epochs of \chandra\/ grating observations, separated by 18 years. This change in the emission line profile morphology is consistent with increased x-ray absorption by the wind, indicative of a mass-loss rate increase of 40 per cent -- a result that is significant at more than the four sigma level. The corresponding wavelength-dependent line flux increase is consistent with the intrinsic x-ray emission increasing along with the wind mass-loss rate, but the emergent line flux being affected by the increased wind absorption. 

This result is quite surprising as \zpup\/ is not known or expected to have a variable mass-loss rate beyond a few percent associated with its observed stochastic and periodic low-level photometric variability. This certainly suggests that continued optical photometric monitoring and spectral monitoring would be recommended as well perhaps as UV spectroscopy and a reanalysis of archival data. Other O and early B supergiants with wind signatures in their x-ray profiles could also be re-observed to see if they too change on years timescales. 

\section*{Acknowledgements}

The scientific results in this article are based on data retrieved from the \chandra\/ data archive. 
ADU acknowledges support from the Natural Sciences and Engineering Research Council of Canada (NSERC). 
MAL acknowledges support from NASA's Astrophysics Program.
LD thanks the Keck Northeast Astronomy Consortium and 
JW and CM thank the Provost's Office of Swarthmore College for supporting their work. 
The authors wish to thank Graham Doskoch for checking some of our initial results and Jon Sundqvist and Alex Fullerton for fruitful discussions about wind and interior properties and H$\alpha$ emission variability. 
And we wish to thank for the referee for comments and suggestions that significantly improved the manuscript.

\section*{Data Availability}

There are no new data generated or analyzed in support of this research. 

\bibliographystyle{mn2e}
\bibliography{zpup}


\appendix

\section{Continuum Regions} 
\label{app:continuum} 

\begin{figure*}
\centering
    \includegraphics[angle=0,width=0.245\textwidth]{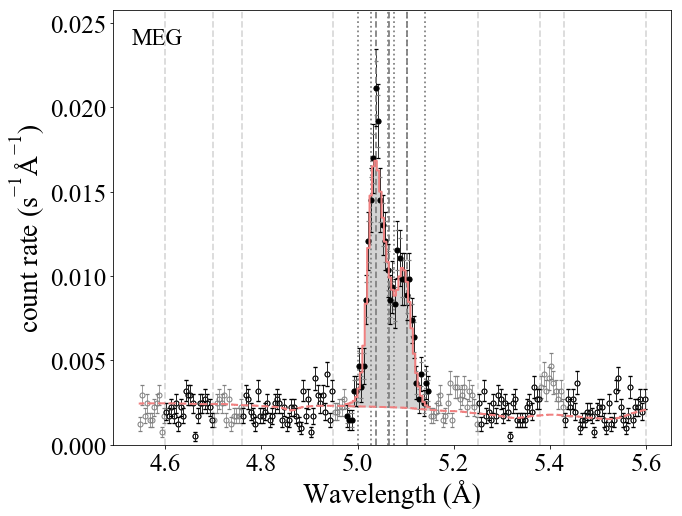}
    \includegraphics[angle=0,width=0.245\textwidth]{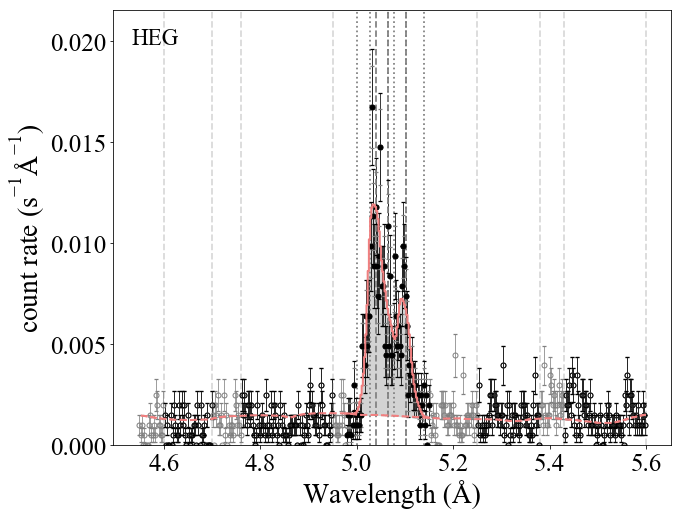}
    \includegraphics[angle=0,width=0.245\textwidth]{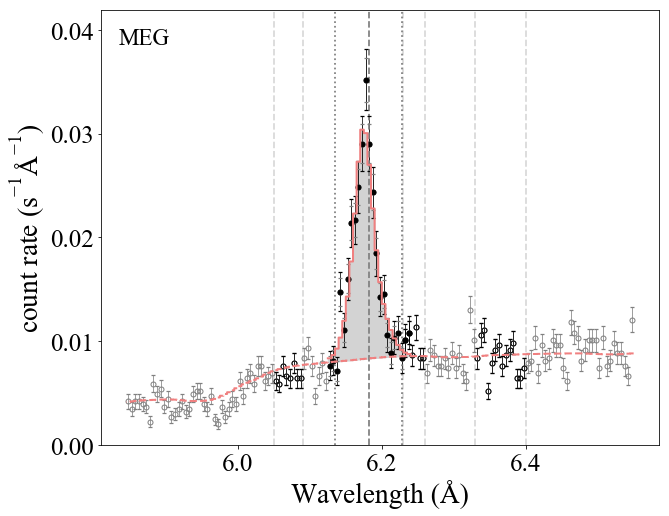}
    \includegraphics[angle=0,width=0.245\textwidth]{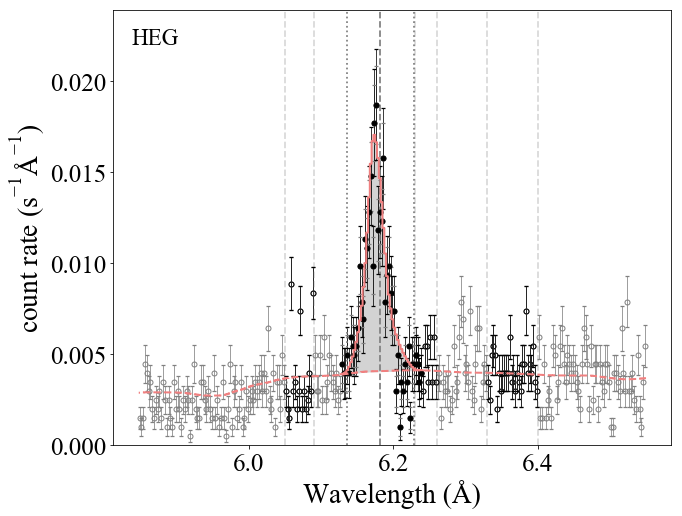}
    
    \includegraphics[angle=0,width=0.245\textwidth]{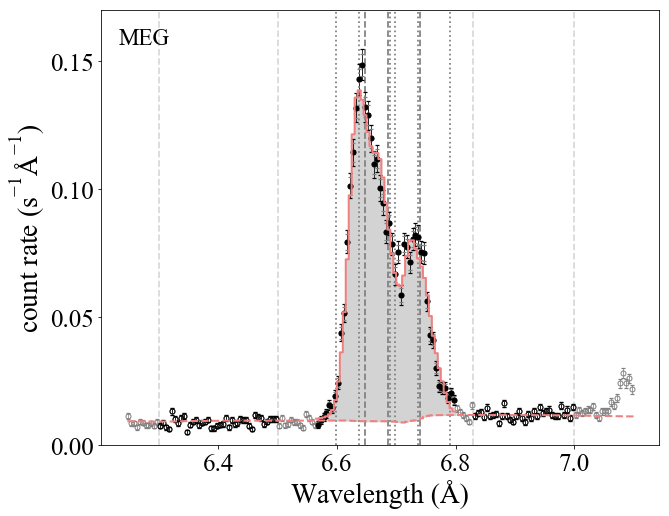}
    \includegraphics[angle=0,width=0.245\textwidth]{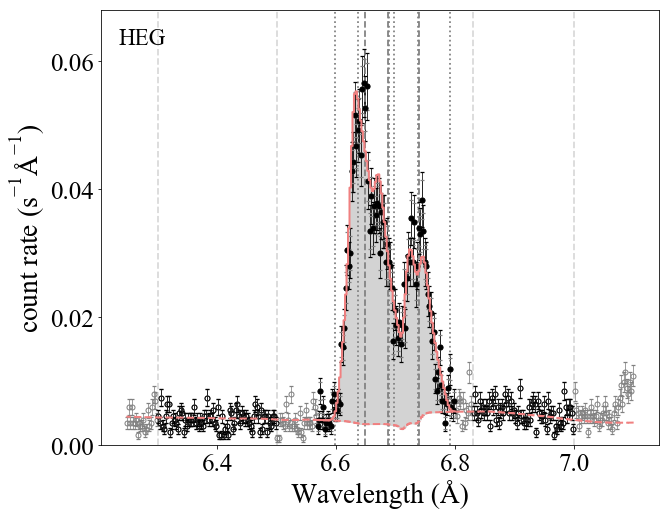}
    \includegraphics[angle=0,width=0.245\textwidth]{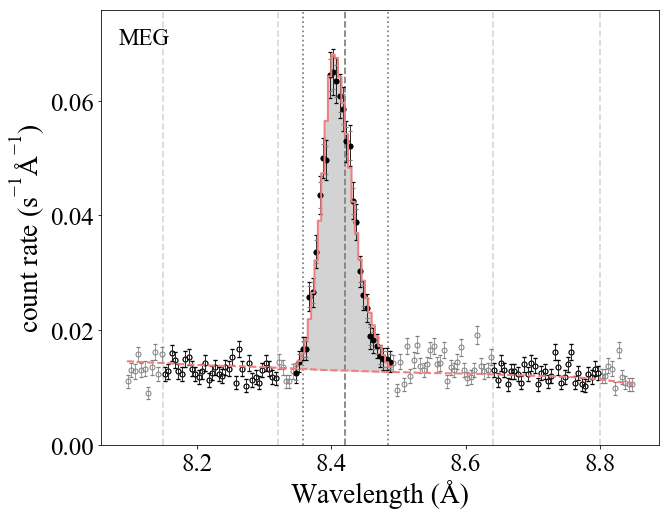}
    \includegraphics[angle=0,width=0.245\textwidth]{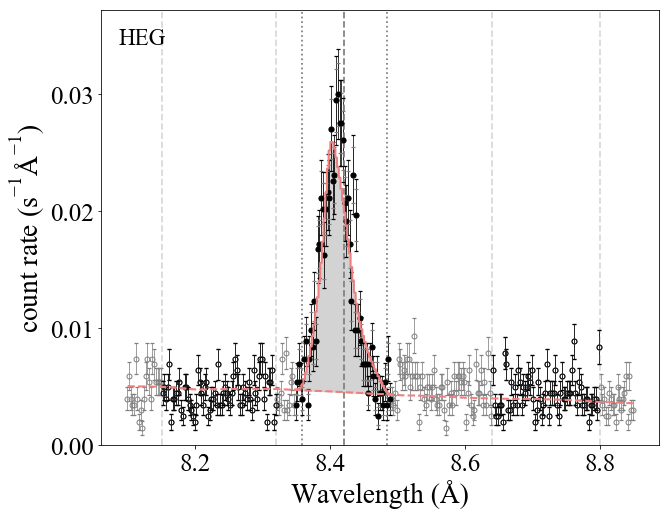}
    
    \includegraphics[angle=0,width=0.245\textwidth]{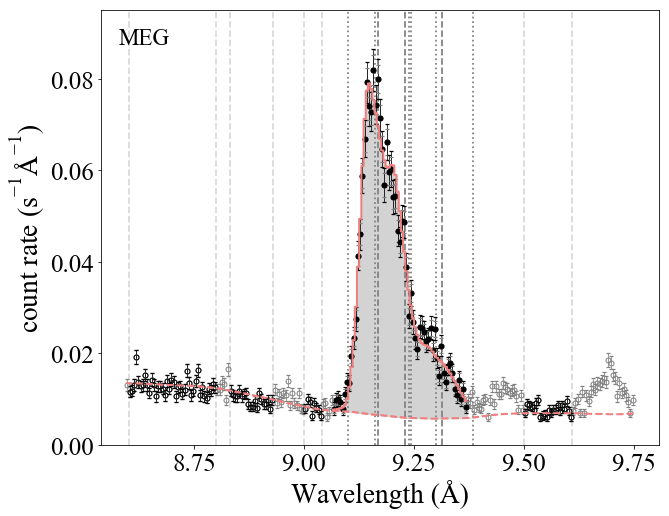}
    \includegraphics[angle=0,width=0.245\textwidth]{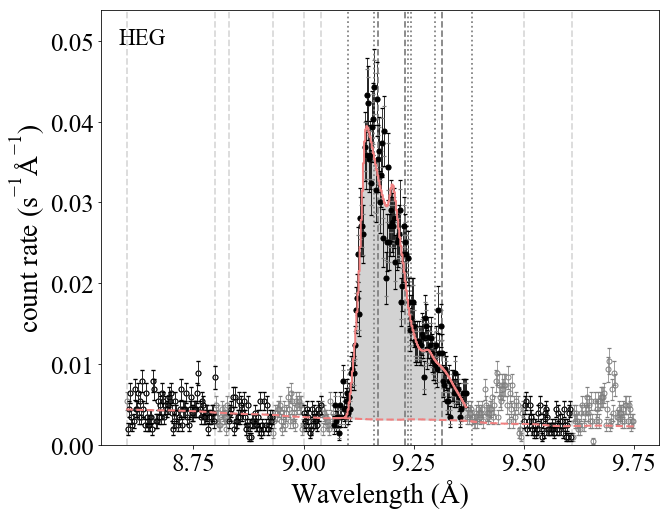}
    \includegraphics[angle=0,width=0.245\textwidth]{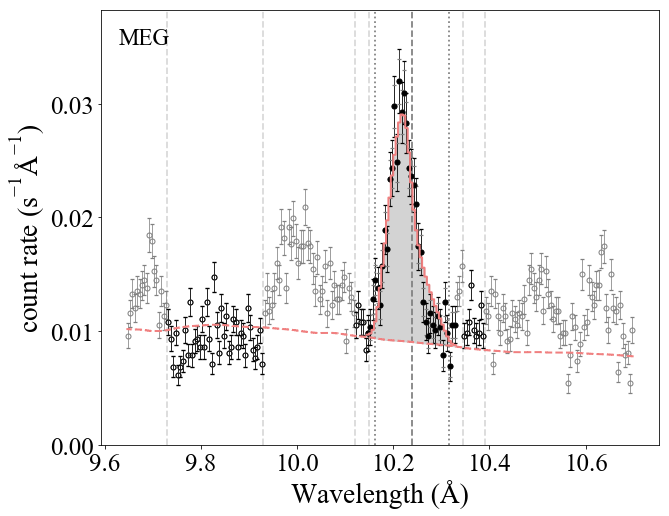}
    \includegraphics[angle=0,width=0.245\textwidth]{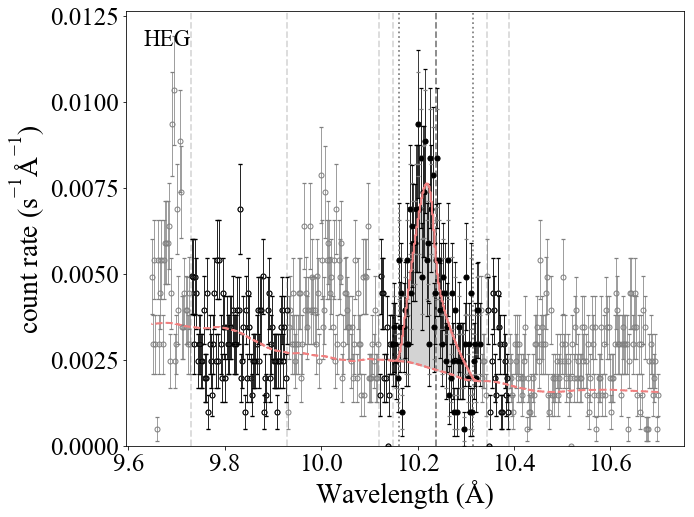}
    
    \includegraphics[angle=0,width=0.245\textwidth]{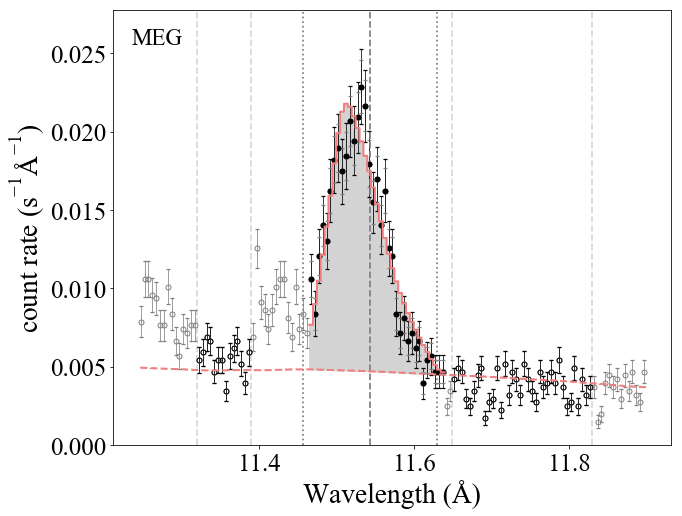}
    \includegraphics[angle=0,width=0.245\textwidth]{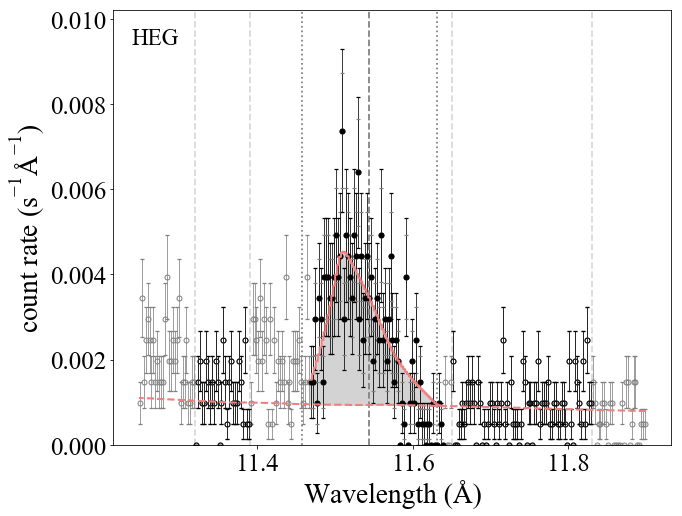}
    \includegraphics[angle=0,width=0.245\textwidth]{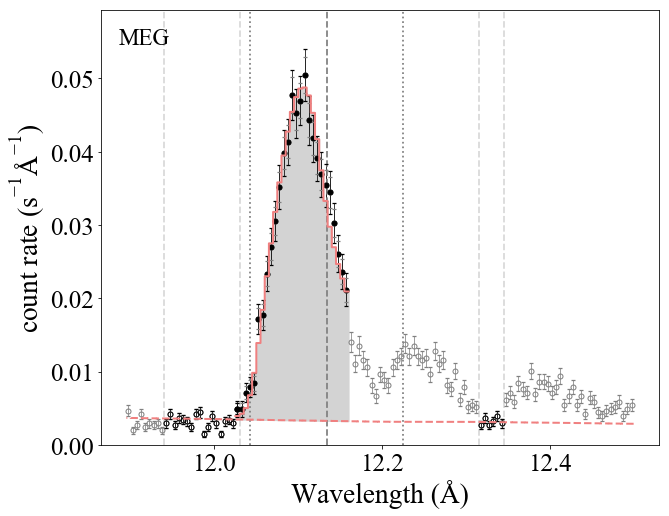}
    \includegraphics[angle=0,width=0.245\textwidth]{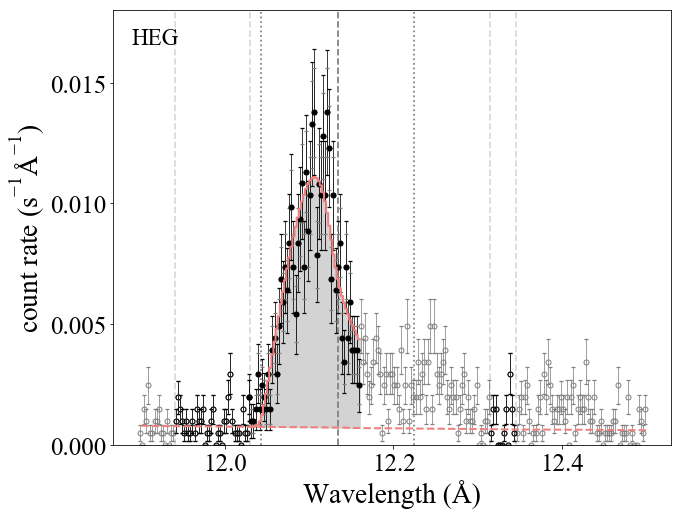}
    
    \includegraphics[angle=0,width=0.245\textwidth]{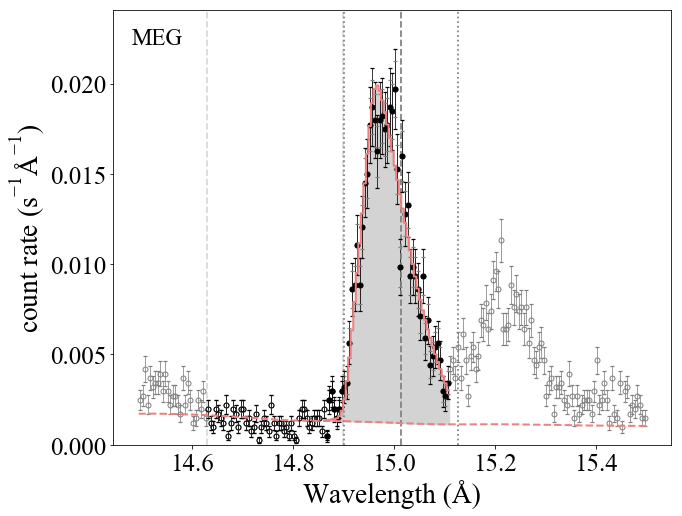}
    \includegraphics[angle=0,width=0.245\textwidth]{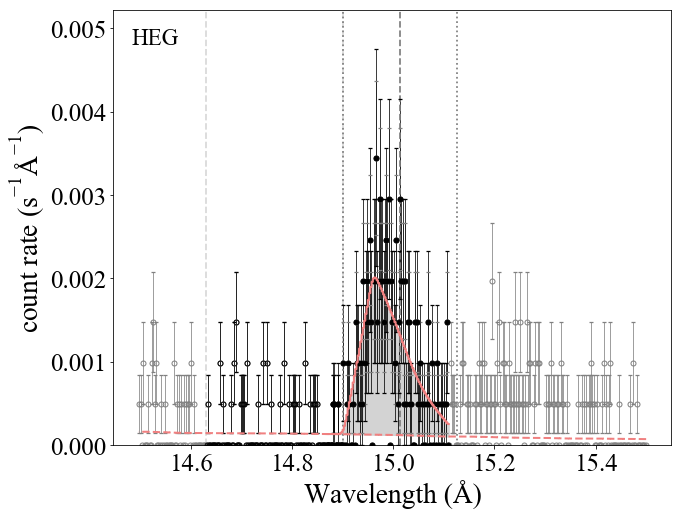}
    \includegraphics[angle=0,width=0.245\textwidth]{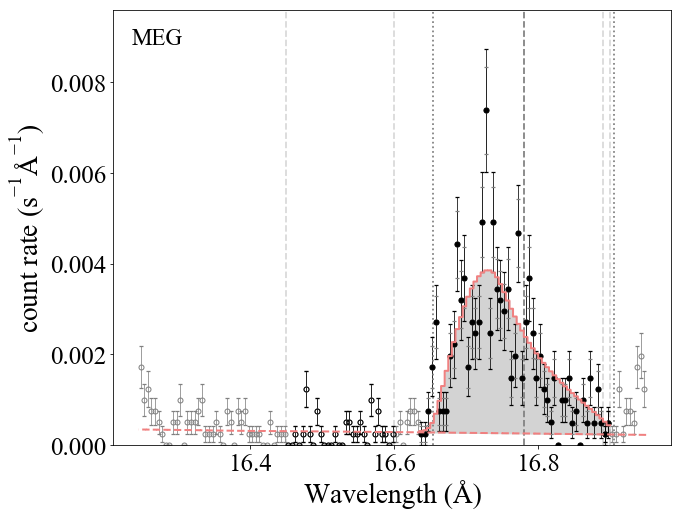}
    \includegraphics[angle=0,width=0.245\textwidth]{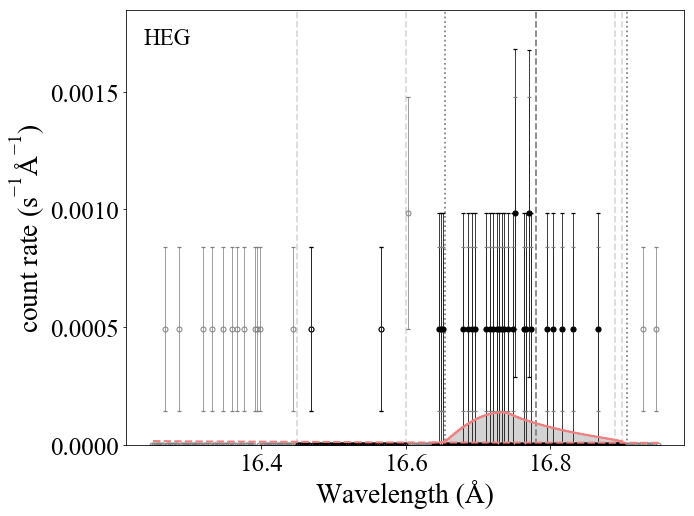}

\caption{Combined 21 cycle 19 Obs IDs are shown for each of the ten lines in MEG and HEG pairs. The pseudo-continuum around each line is indicated, with both vertical dotted lines and darkened data points showing the regions that were used for the fitting. 
    }	

\label{fig:continuum_regions_plots}
\end{figure*}

We carefully evaluated the wavelength regions used for fitting the continuum around each line, both by visually inspecting the data and by consulting the {\sc atomdb} line list \citep{Foster2012}. Of course, a large number of very weak lines are distributed throughout the continuum, making it in reality a pseudo-continuum. The wavelength regions relatively free of contaminating lines are listed in Table \ref{tab:continuum_regions_tabel}. They and the associated continuum fits for the cycle 19 data are shown graphically in Fig.\ \ref{fig:continuum_regions_plots}, along with the line fits themselves.  

\begin{table}
  \caption{Emission line pseudo-continuum regions}
\begin{tabular}{ccc}
  \hline
ion &  line wavelength & continuum ranges \\
    & (\AA)          & (\AA)\\
\hline
S {\sc xv}    & 5.04, 5.07, 5.10  & 4.60-4.70, 4.76-4.95, 5.25-5.38, 5.43-5.60\\
Si {\sc xiv}  & 6.18              & 6.05-6.09, 6.23-6.26, 6.33-6.40\\
Si {\sc xiii} & 6.65, 6.69, 6.74  & 6.30-6.50, 6.83-7.00\\
Mg {\sc xii}  & 8.42              & 8.15-8.32, 8.64-8.80\\
Mg {\sc xi}   & 9.17, 9.23, 9.31  & 8.60-8.80, 8.83-8.93, 9.00-9.04, 9.50-9.61\\
Ne {\sc x}    & 10.24             & 9.73-9.93, 10.12-10.15, 10.345-10.39\\
Ne {\sc ix}   & 11.54             & 11.32-11.39, 11.65-11.83\\
Ne {\sc x}    & 12.13             & 11.94-12.03, 12.315-12.345\\
Fe {\sc xvii} & 15.01             & 14.63-14.90\\
Fe {\sc xvii} & 16.78             & 16.45-16.60, 16.89-16.90\\
\hline
\end{tabular}
\label{tab:continuum_regions_tabel}
\end{table}  

\section{Wind profile fitting of individual lines} 
\label{app:profiles} 

\begin{figure*}
\centering

    \includegraphics[angle=0,width=0.245\textwidth]{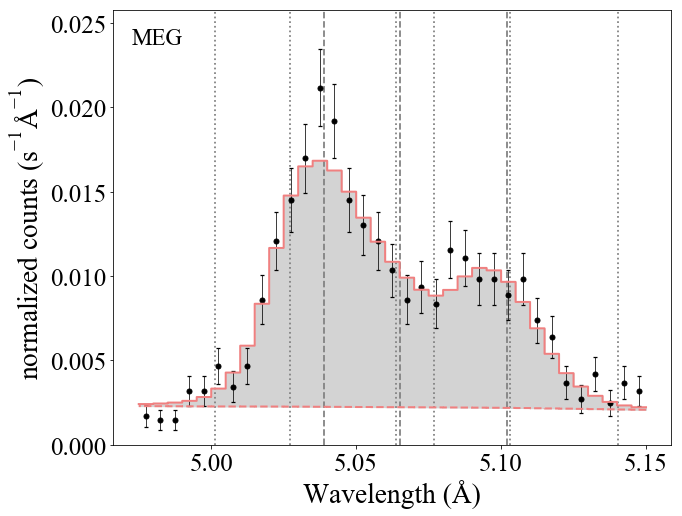}
    \includegraphics[angle=0,width=0.245\textwidth]{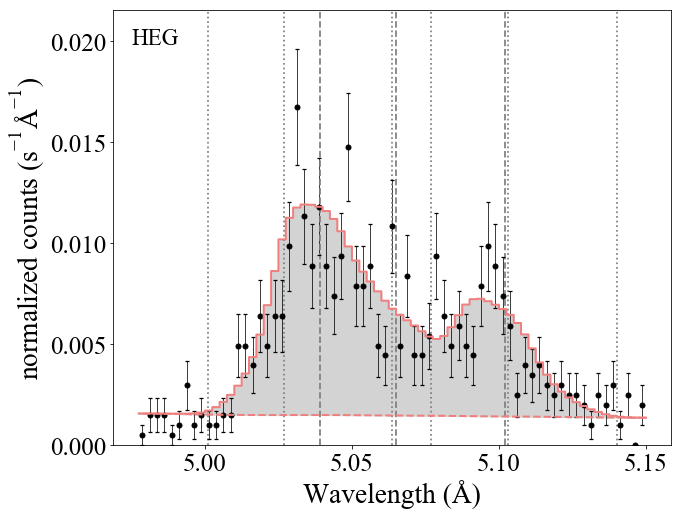}
    \includegraphics[angle=0,width=0.245\textwidth]{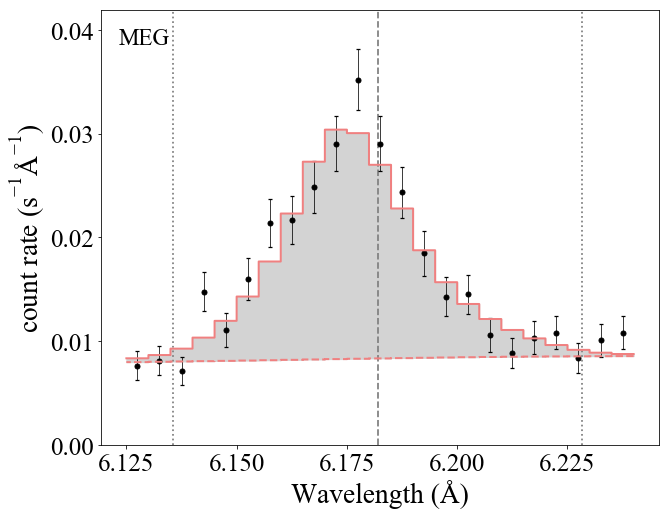}
    \includegraphics[angle=0,width=0.245\textwidth]{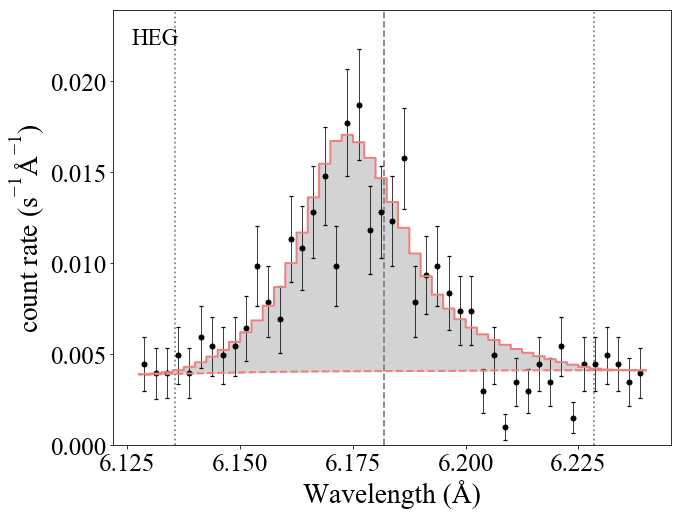}

    \includegraphics[angle=0,width=0.245\textwidth]{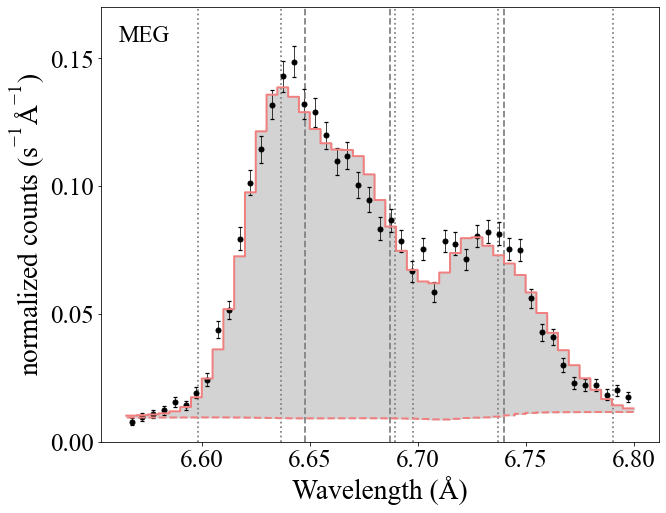}
    \includegraphics[angle=0,width=0.245\textwidth]{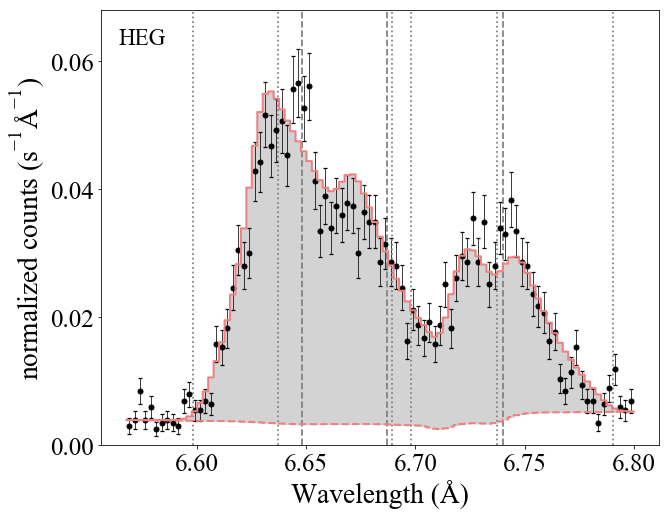}
    \includegraphics[angle=0,width=0.245\textwidth]{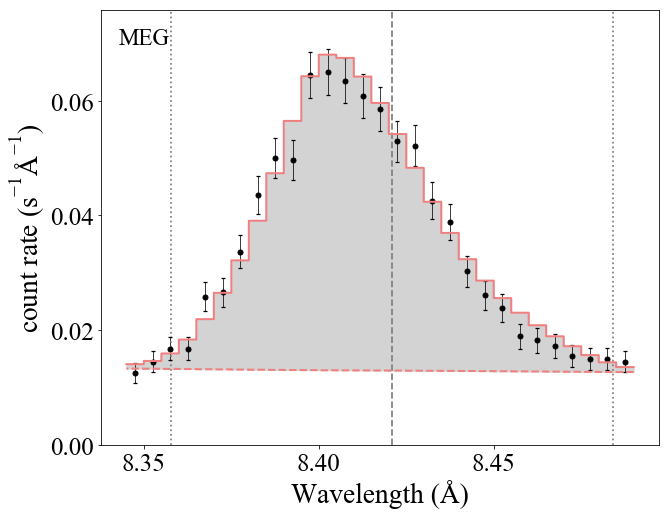}
    \includegraphics[angle=0,width=0.245\textwidth]{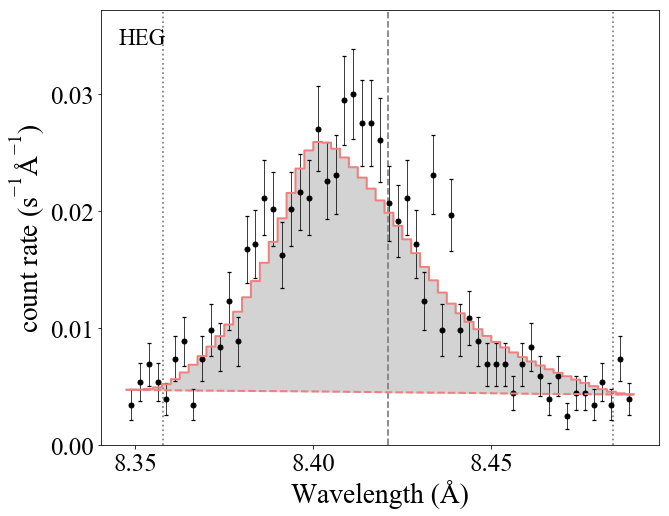}

    \includegraphics[angle=0,width=0.245\textwidth]{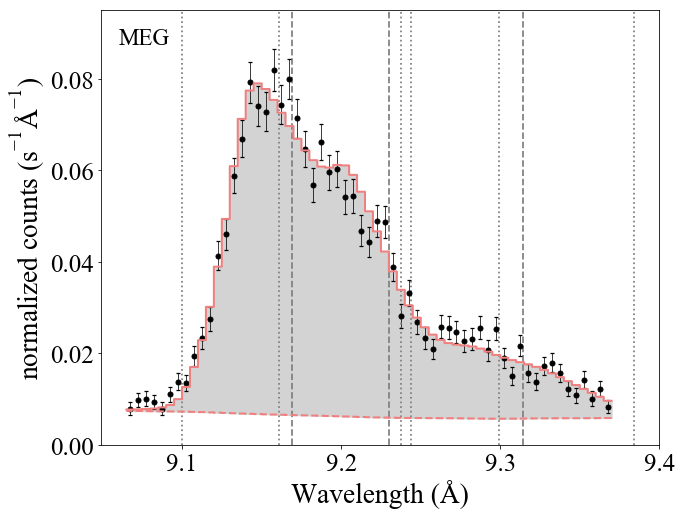}
    \includegraphics[angle=0,width=0.245\textwidth]{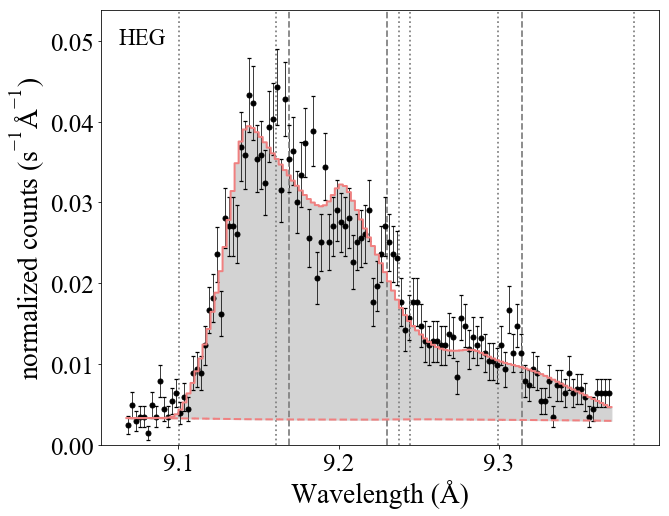}
    \includegraphics[angle=0,width=0.245\textwidth]{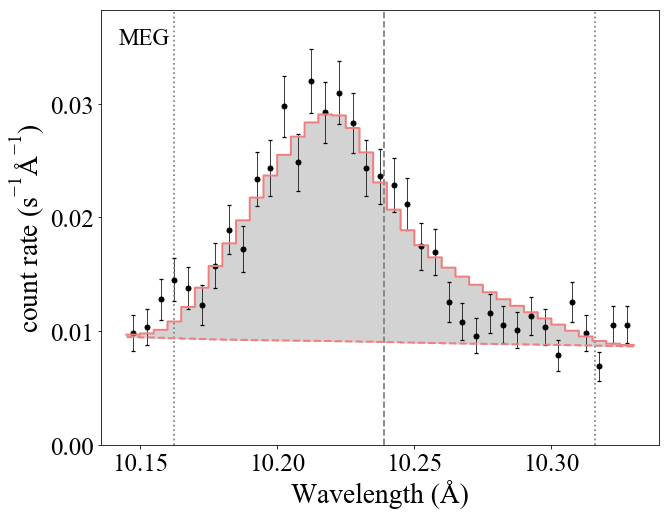}
    \includegraphics[angle=0,width=0.245\textwidth]{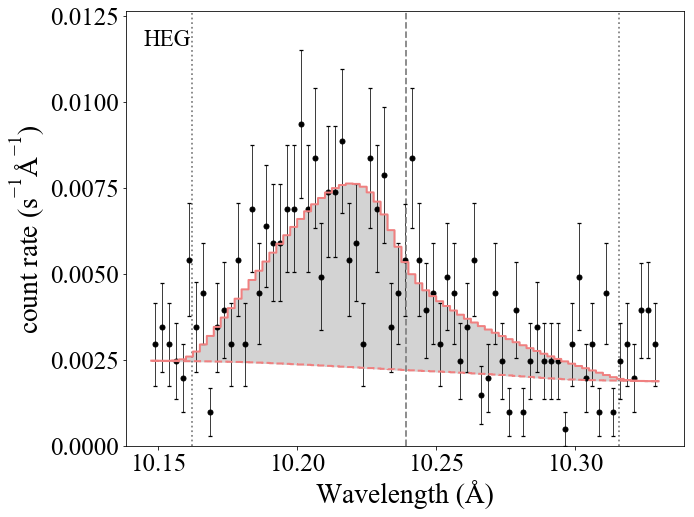}

    \includegraphics[angle=0,width=0.245\textwidth]{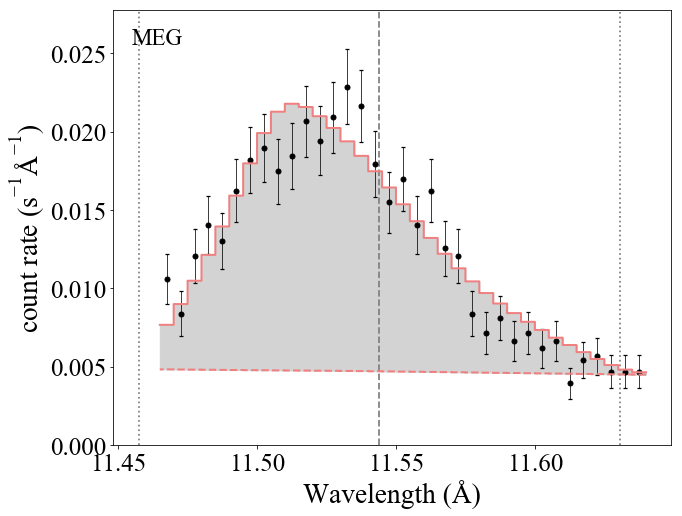}
    \includegraphics[angle=0,width=0.245\textwidth]{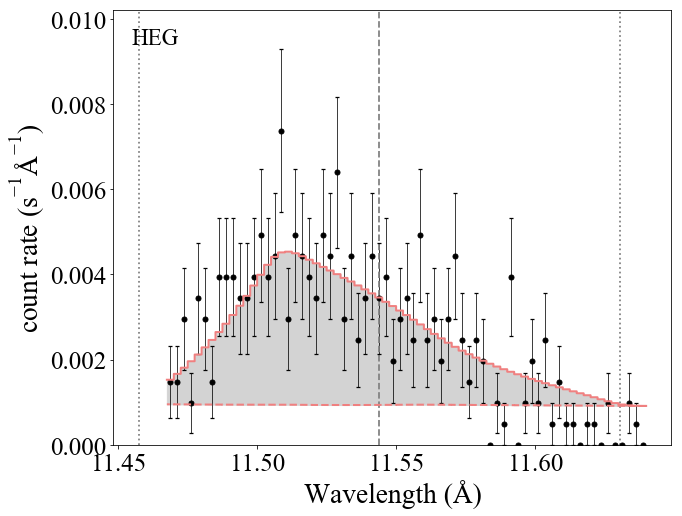}
    \includegraphics[angle=0,width=0.245\textwidth]{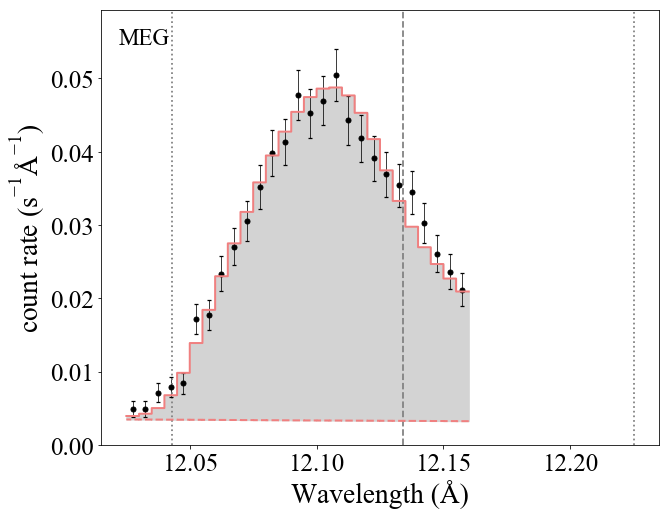}
    \includegraphics[angle=0,width=0.245\textwidth]{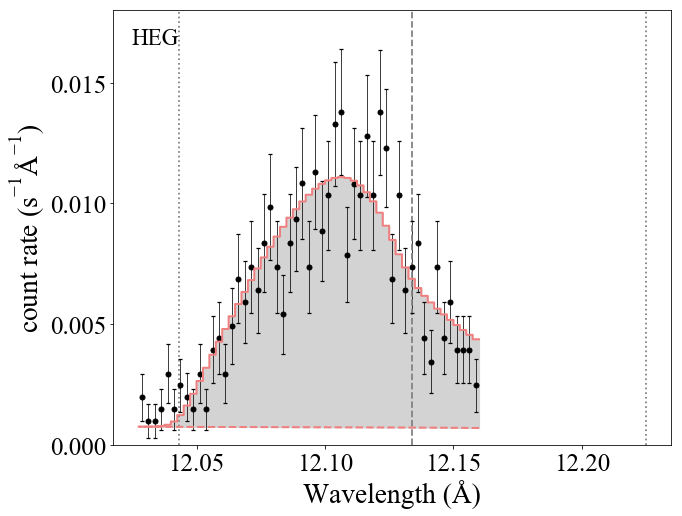}

    \includegraphics[angle=0,width=0.245\textwidth]{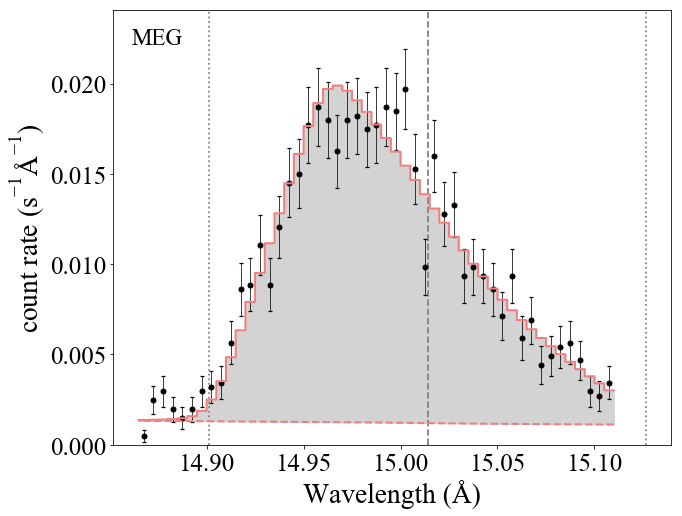}
    \includegraphics[angle=0,width=0.245\textwidth]{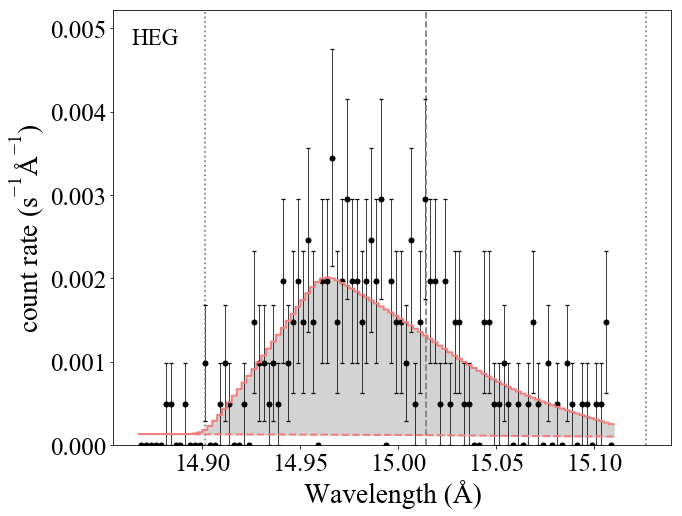}
    \includegraphics[angle=0,width=0.245\textwidth]{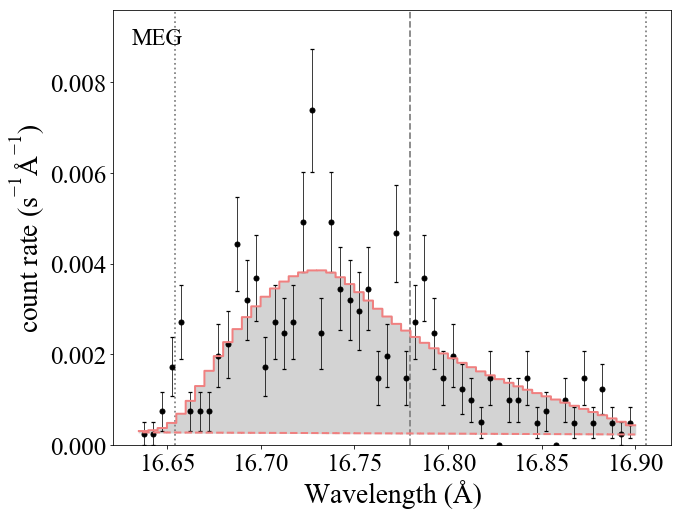}
    \includegraphics[angle=0,width=0.245\textwidth]{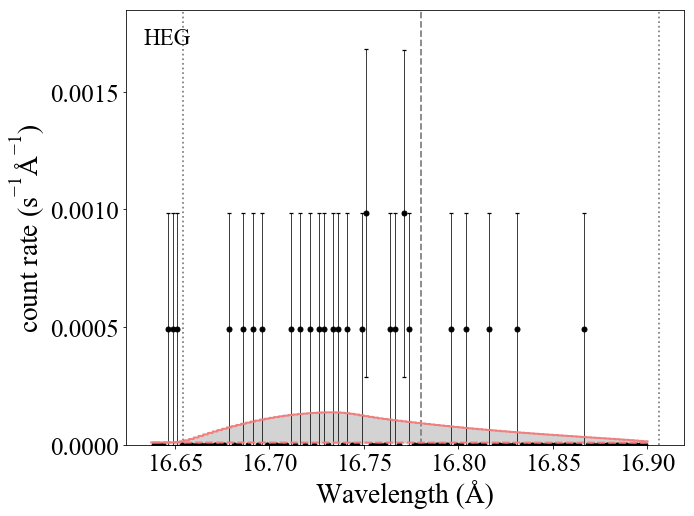}
    
\caption{We show wind profile models (on top of continuum models) fit to the cycle 19 data for each of the ten lines and line complexes analyzed in this paper. For each line or complex we show a pair of panels: combined first order MEG (left; so first and third columns) and HEG (right; so second and fourth columns). A darker vertical dashed line indicates the rest wavelength of each line, while lighter flanking vertical lines indicate the Doppler shifted wavelengths associated with the wind terminal velocity of $2250$ km s$^{-1}$.
 	}
\label{fig:line_profiles}
\end{figure*}

\begin{figure*}
\centering
    \includegraphics[angle=0,width=0.245\textwidth]{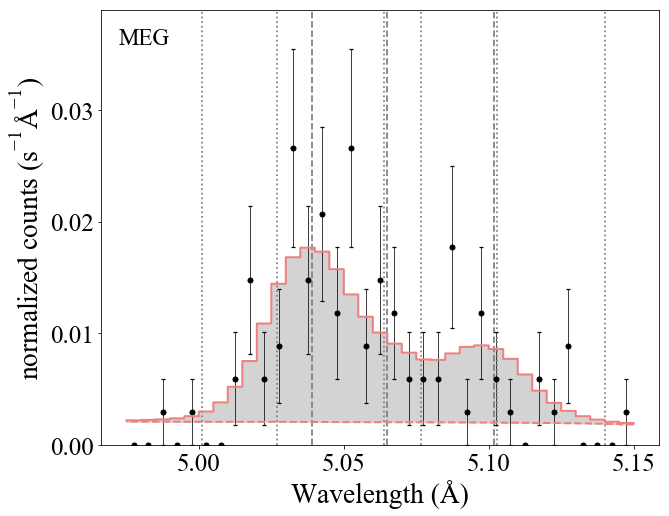}
    \includegraphics[angle=0,width=0.245\textwidth]{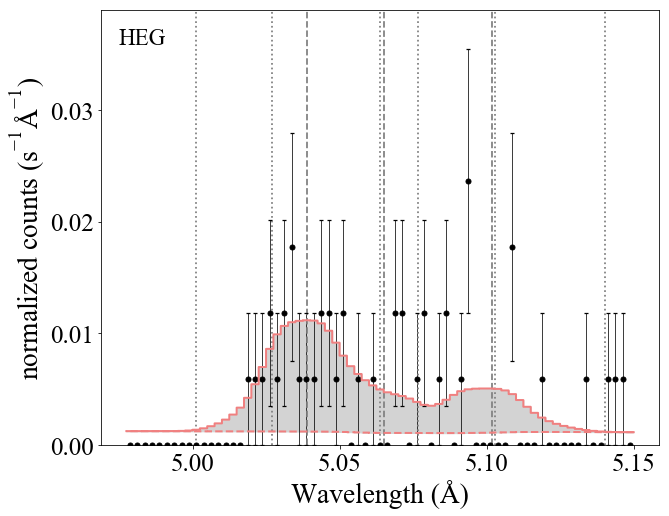}
    \includegraphics[angle=0,width=0.245\textwidth]{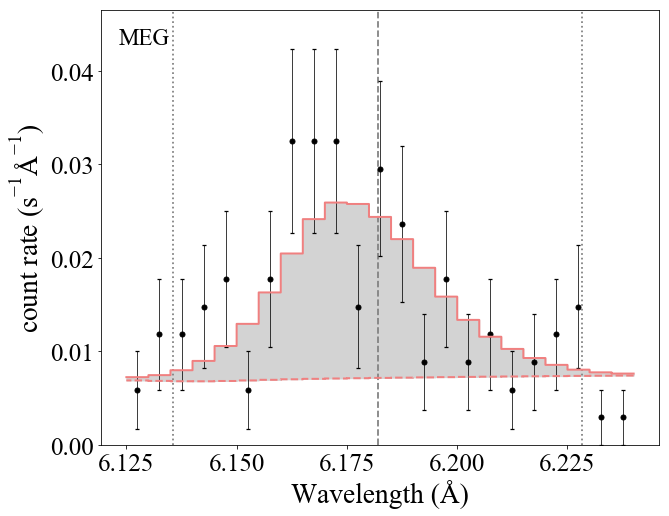}
    \includegraphics[angle=0,width=0.245\textwidth]{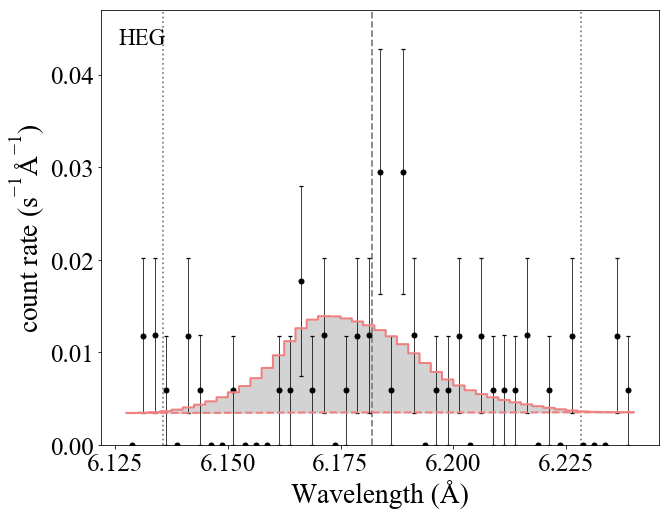}
    
    \includegraphics[angle=0,width=0.245\textwidth]{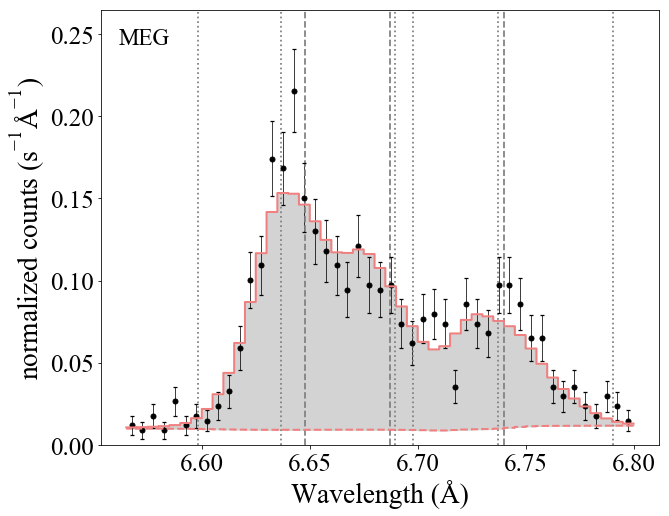}
    \includegraphics[angle=0,width=0.245\textwidth]{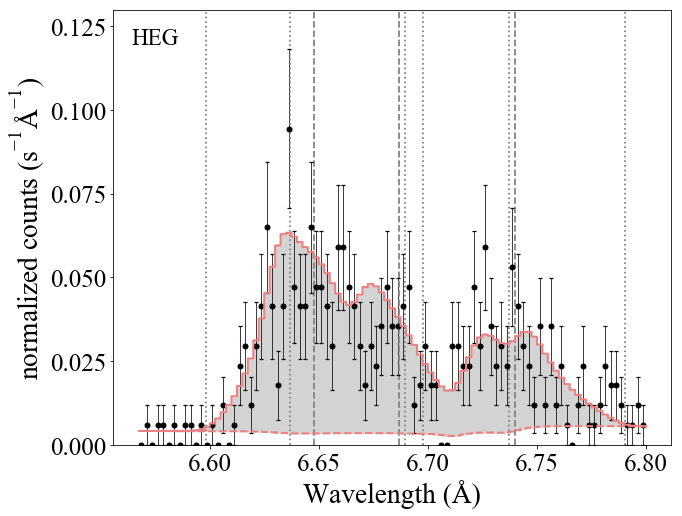}
    \includegraphics[angle=0,width=0.245\textwidth]{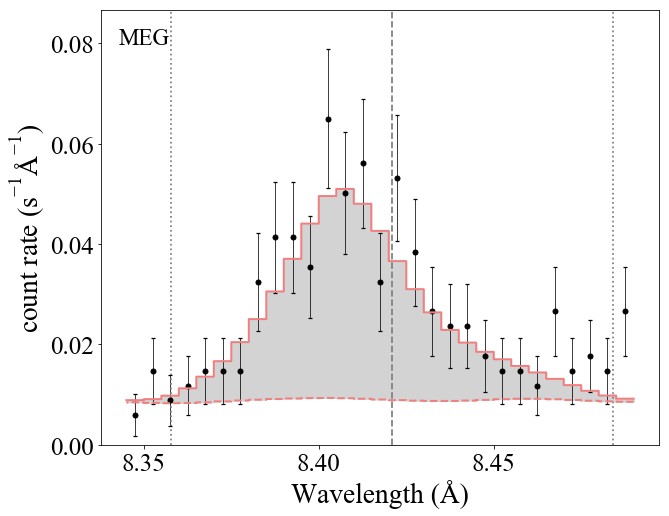}
    \includegraphics[angle=0,width=0.245\textwidth]{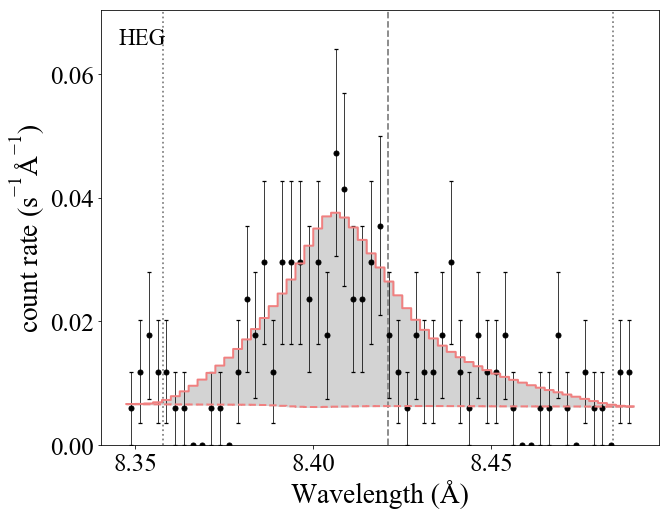}
    
    \includegraphics[angle=0,width=0.245\textwidth]{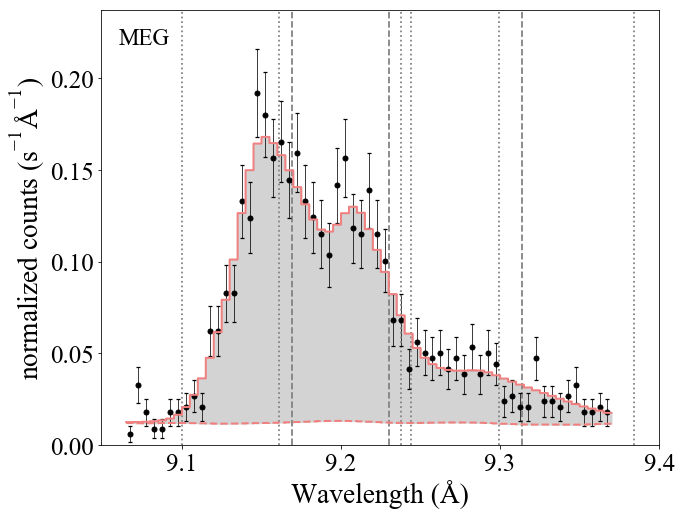}
    \includegraphics[angle=0,width=0.245\textwidth]{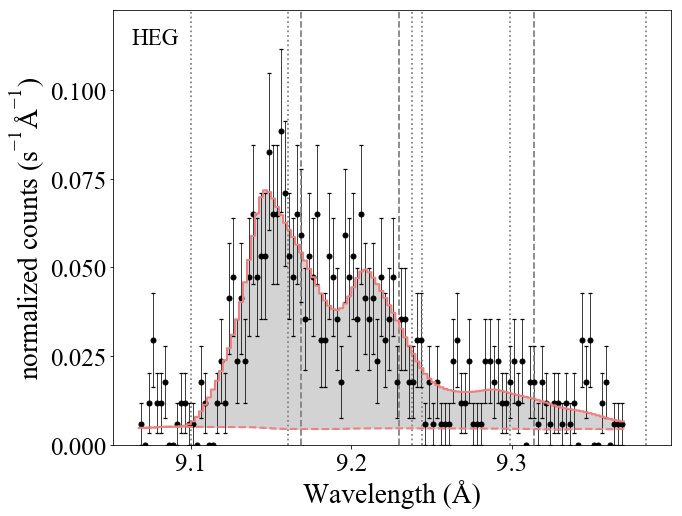}
    \includegraphics[angle=0,width=0.245\textwidth]{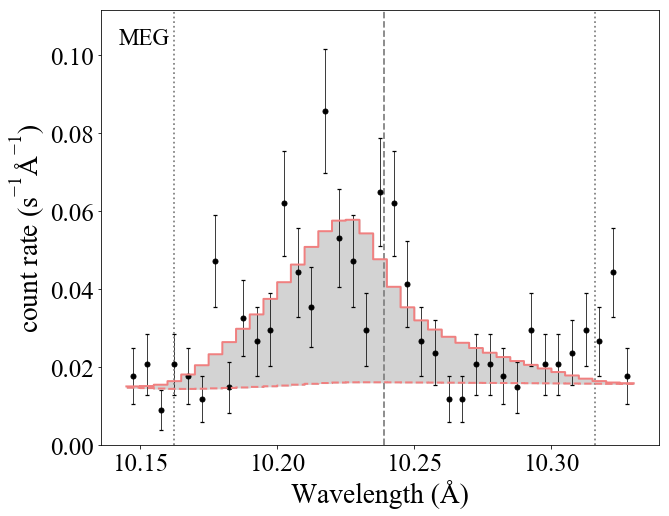}
    \includegraphics[angle=0,width=0.245\textwidth]{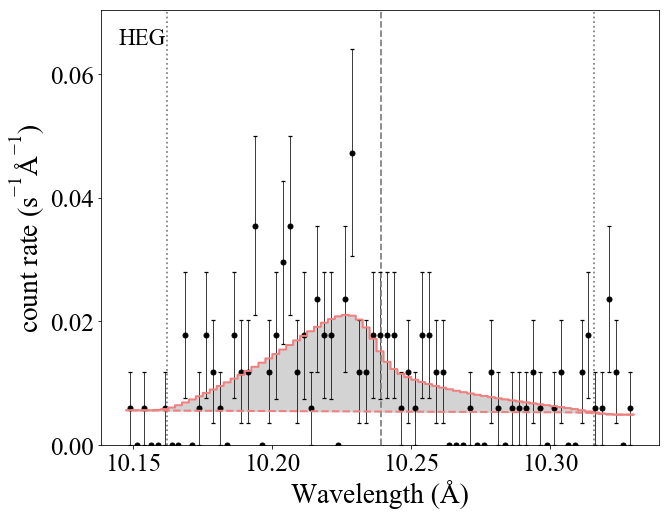}
    
    \includegraphics[angle=0,width=0.245\textwidth]{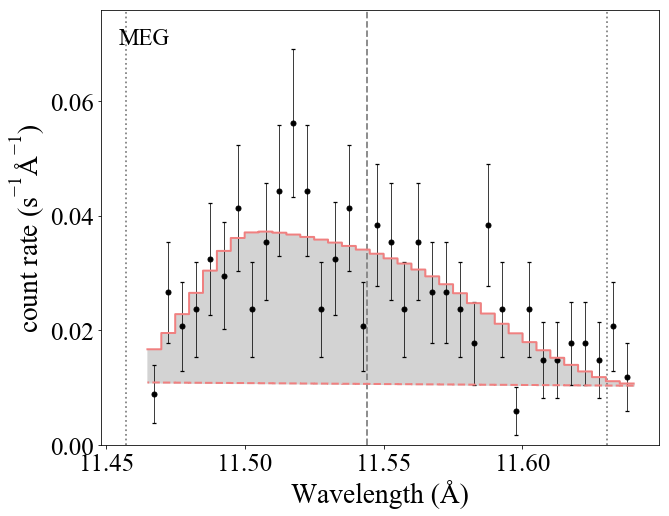}
    \includegraphics[angle=0,width=0.245\textwidth]{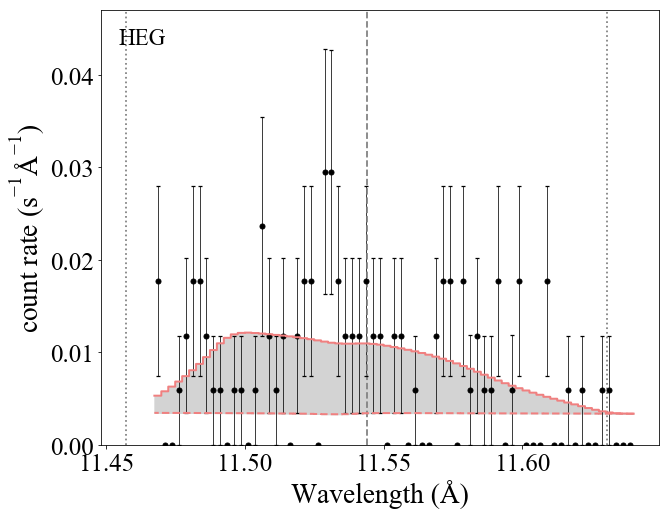}
    \includegraphics[angle=0,width=0.245\textwidth]{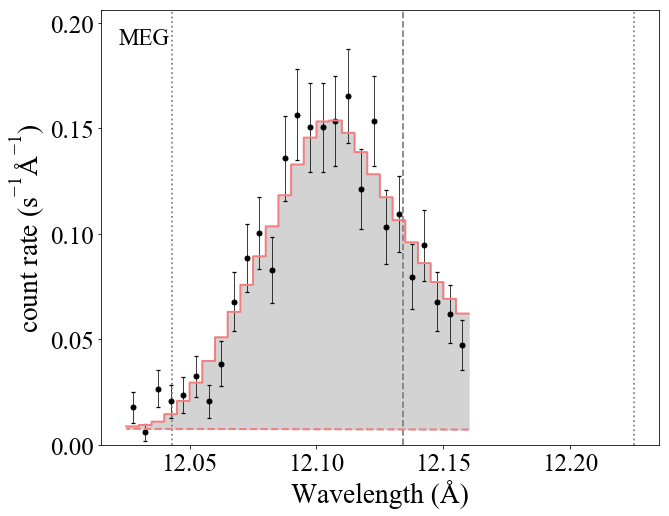}
    \includegraphics[angle=0,width=0.245\textwidth]{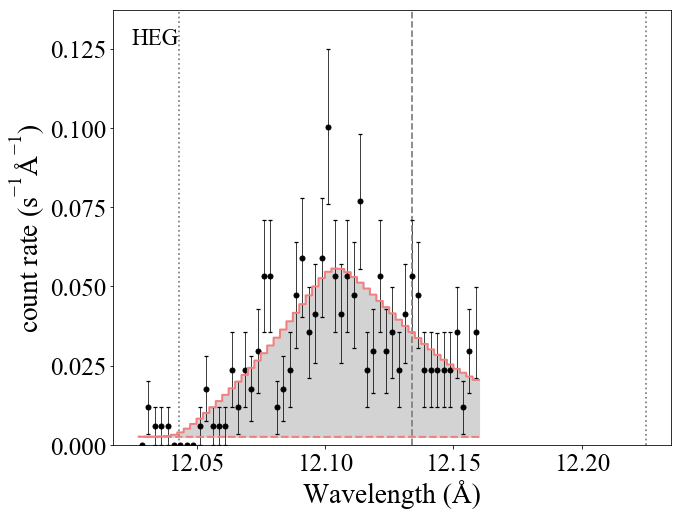}
    
    \includegraphics[angle=0,width=0.245\textwidth]{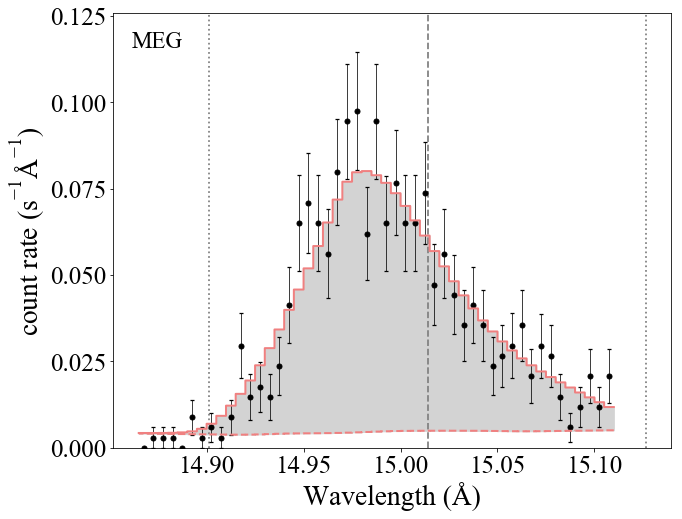}
    \includegraphics[angle=0,width=0.245\textwidth]{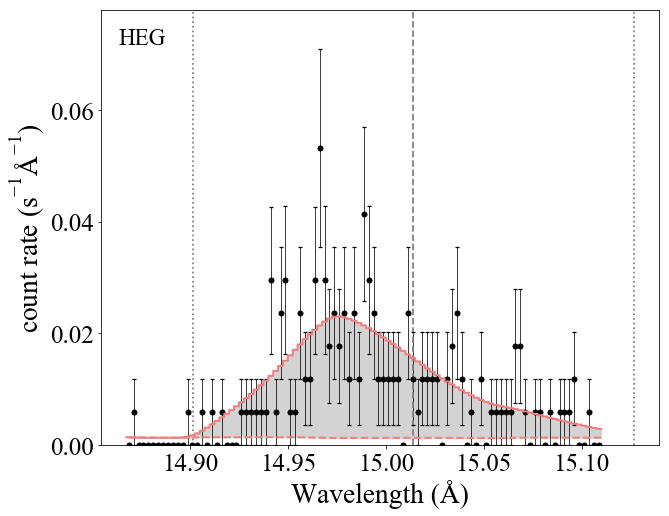}
    \includegraphics[angle=0,width=0.245\textwidth]{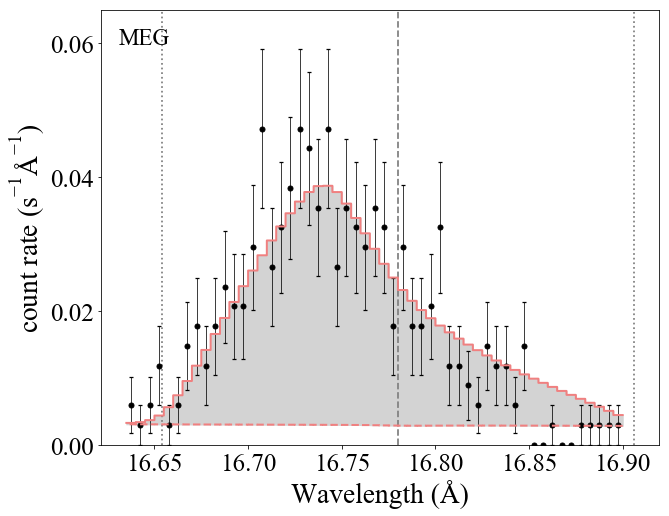}
    \includegraphics[angle=0,width=0.245\textwidth]{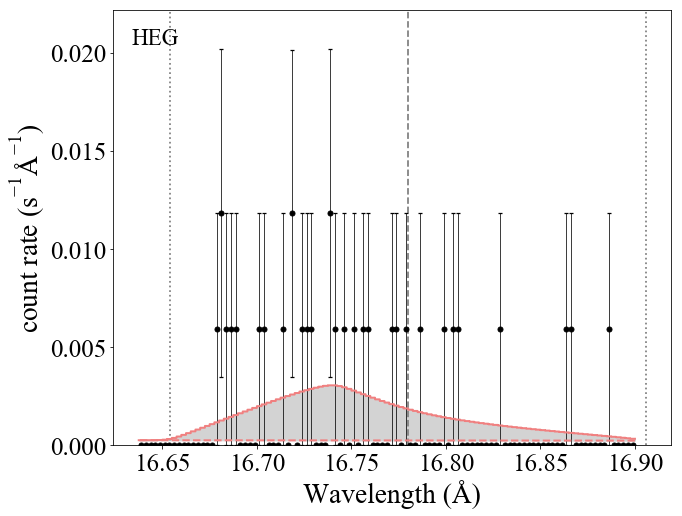}
    
	\caption{Same as Fig.\ \ref{fig:line_profiles} but for cycle 1. 
	}	
\label{fig:640_line_profiles}
\end{figure*}

The fits to each line and line complex are shown in Fig.\ \ref{fig:line_profiles} for the cycle 19 data and in Fig.\ \ref{fig:640_line_profiles} for the cycle 1 data. 

\end{document}